\documentclass[twocolumn,showpacs,showkeys,preprintnumbers,amsmath,amssymb]{revtex4}

\usepackage{graphicx}
\usepackage{dcolumn}
\usepackage{bm}
\usepackage{amsmath}
\usepackage{tabularx}
\usepackage{graphics}
\usepackage{epsfig}
\usepackage{amssymb}
\usepackage{multirow}
\newcolumntype{R}[1]{>{\raggedleft\arraybackslash}p{#1}}

\begin{document}
\vspace*{2.0 cm}



 \title{Detailed description of exclusive muon capture rates using realistic two-body forces}



\author{ P.G. Giannaka }
\email{ pgiannak@cc.uoi.gr}
\author{ T.S Kosmas }
\email{hkosmas@uoi.gr}
\affiliation{Division of Theoretical Physics, University of Ioannina, GR-45110 Ioannina, Greece }

\date{\today}

\begin{abstract}
Starting from state-by-state calculations of exclusive rates of the ordinary muon capture (OMC), we evaluated total $\mu^{-}$-capture rates for a set of light- and medium-weight nuclear isotopes. We employed a version of the proton-neutron quasi-particle random phase approximation (pn-QRPA, for short) which uses as realistic nuclear forces the Bonn C-D one boson exchange potential. Special attention was paid on the percentage contribution to the total $\mu^{-}$-capture rate of specific low-spin multipolarities resulting by summing over the corresponding multipole transitions. The nuclear method used offers the possibility of estimating seperately the individual contributions to the total and partial rates of the polar-vector and axial-vector components of the weak interaction Hamiltonian for each accessible final state of the daughter nucleus. One of our main goals is to provide a reliable description of the charge changing transitions matrix elements entering the description of other similar semileptonic nuclear processes like the charged-current neutrino-nucleus reactions, the electron capture on nuclei, the single $\beta^{\pm}$-decay mode, etc., which play important role in currently interesting laboratory and astrophysical applications like the  neutrino-detection through lepton-nucleus interaction probes, and neutrino-nucleosynthesis. Such results can be also be useful in various ongoing muon-capture experiments at PSI, Fermilab, JPARC and RCNP. 
\end{abstract}


\maketitle

\section{Introduction} 
\label{Introduction}
In recent years, various sensitive experiments take advantage of the powerful muon beams produced in well known muon factories (PSI, Fermilab, JPARC, RCNP in Osaka and others) for standard and non-standard muon physics probes \cite{Kuno-2013,Ejiri}. Among the standard model probes those involving muon capture on nuclei specifically those emitting X-rays and/or several particles (p, n, $\alpha$, etc.) after $\mu^-$-capture, which are important to understand the rates and spectra of these particles, are investigated \cite{Kuno-2013}. For example, at PSI researchers are interested in experiments based on the emission of charged particles from muonic atoms of Al, Si and Ti or neutron emission following muon capture from Fe, Ca ,Si and Al \cite{Kuno-2013}. Also very recently, in the highly intense muon facilities MuSIC at RCNP, Osaka, Japan, nuclear muon capture reactions (on Mo, Pb, etc.) are planned in order to study, nuclear weak responses (for neutrino reactions, etc.) \cite{Ejiri}. For experiments like the above, it is important, before going to the rates of the emitted particles, to know the first stage muon capture process. 


As it is well known, when negative muons, $\mu^{-}$, produced in a meson factory, slow down in matter, it is possible for them to be captured in atomic orbits. Afterwards, fast electromagnetic cascades bring these muons down to the innermost (1s or 2p) quantum orbits (in this way muonic atoms are produced) \cite{Con-Don-72,DonPe,Mukh-77,Measd-01,Suz-Maes-87}. A bound muon in the muonic atom may disappear either by decay known as muon decay in orbit or by capture by the nucleus the main channel of which is the ordinary muon capture represented by the reaction \cite{Kol-Lang-94,Kol-Lang-Vog-97,Kol-Lang-00}
\begin{equation}
 \mu_{b}^{-} + (A,Z) \rightarrow (A,Z-1)^{\ast} + \nu_{\mu}, 
\label{mucap}
\end{equation}
where $(A,Z)$ denotes the initial atomic nucleus with mass number $A$ and proton number $Z$ while $(A,Z-1)^{\ast}$ stands for an excited state of the daughter nuclear isotope.

The reaction (\ref{mucap}) is a well known example of symbiosis of atomic, nuclear and particle physics.  
In this work, however, we will concentrate on its nuclear physics aspects. As muon-capture in nuclei presents many advantages for the study of both nuclear structure and the fundamental electro-weak interactions \cite{Borie-Rinker-82,Kosmas-01,Don-Wal-72,Don-Wal-73}, process (\ref{mucap}) has been the subject of extensive 
experimental and theoretical investigations started early on 50's  using closure approximation or sum-over partial rates to find the total $\mu^-$-capture rate (the measured quantity).
\cite{Mukh-77,Rose-77,Rose-78,Measd-01,Foldy-Wal-64,Giai-Auerb-81,Con-Don-72,DonPe,Auer-Zam-82,Auer-Kle-84,Urin-Vyaz-92}.
In the plethora of the relevant papers, the most important motivation
was rested on the hope to explain how nucleons (hadronic current) inside the nucleus couple weakly to 
the lepton field (leptonic current). 
The nuclear physics aspects of process (\ref{mucap}), however, still possess some yet unresolved fundamental problems, e.g. those related to the  nucleon-nucleon and lepton-nucleus interactions, the question whether the individual properties of the nucleons change when they are packed together 
in the nucleus or remain essentially unaffected like the coupling of the nucleon to the leptonic field, etc. 

The interest of studying $\mu^{-}$-capture has recently been revived \cite{Zin-Lang-06,Eram-Kuz-Tet-98,Kuz-Tet-02}
due to its prominent role in testing the nuclear models employed in several physical applications in neutrino physics and astrophysics \cite{Lang-Pin-03,Dean-Lang-98,Kol-Lang-99}.
Specifically, $\mu^{-}$-capture is a very useful test for various nuclear methods used to describe semi-leptonic
weak charged-current reactions \cite{Foldy-Wal-64,Kol-Lang-94} as the electron capture in stars (critical in the collapse 
of supernovae) \cite{Lang-Pin-03,Dean-Lang-98,Giannaka}, the neutrino nucleus scattering (important in the detection of astrophysical neutrinos) 
\cite{Lang-Pin-03,Kol-Lang-99}, and other reactions \cite{Kosmas-01}. This is due to the fact that the muon-capture involves a large 
momentum transfer and, hence, it can provide valuable information about effects which are not found in processes like the beta-decay modes 
(on medium momentum transfer processes however, useful information can be also obtained  from low-spin forbidden transitions of beta-decays and charge exchange reactions) \cite{Must-Suh-07}.
Furthermore, there is an intimate relation between the inclusive muon capture rate and the cross section for the 
antineutrino induced charged-current reactions, since both are governed by the same nuclear matrix 
elements and proceed from the same set of initial to the same final nuclear states \cite{Lang-Pin-03,Dean-Lang-98,Kol-Lang-99,Kol-Lang-00}.
Moreover, from the ground state transition matrix elements of the $\mu^-$-capture process one may also derive cross sections for the beta decay modes \cite{Giannaka}.
Calculations on single beta decay which are more difficult to calculate, need explicit nuclear structure calculations \cite{Gian-Kosm-15}.

The purpose of the present work is to perform detailed state-by-state calculations 
\cite{Kos-Verd-94,Kos-Faes-97,Has4,Ts-Kos-11,Bal-Ydr-11,Gian-Kos-13,Balasi-Ydr-11,tsak-kos-11,Bal-Ydr-12,kos-ts-12}
of exclusive muon capture rates and concentrate on the individual contribution of each basic multipole operator 
inducing low-lying excitations in the daughter nucleus. In contrast, most of the previous muon capture
calculations have been performed within the assumptions of closure approximation \cite{Rose-77,Rose-78,Primak-59}. Towards 
this aim, the pn-QRPA provides a reliable description of the required nuclear transition matrix elements 
\cite{Don-Wal-72,Don-Wal-73,Con-Don-72,Don-Wal-76,DonPe,Bar-60,Kam-Faes-91,Rod-Faes-06,Yous-Faes-09,Sam-Krm-10,Nabi-Klap-99}.
Our extensive channel-by-channel calculations would be carried out  
for the exclusive, partial and total muon capture rates, and the results refer to the nuclear isotopes $^{28}Si, ^{32}S, ^{48}Ti, ^{56}Fe, ^{66}Zn $ and $^{90}Zr$, which cover the light- and medium-weight region of the periodic table. We also specialize on the individual contributions of the   polar-vector and axial-vector components of the $\mu^{-}$-capture operators in each of the multipole states and in the total Ordinary Muon Carture (OMC) rate.
Despite the fact that the semileptonic process (\ref{mucap}) is studied for a long time \cite{Mukh-77,Suz-Maes-87,Measd-01,Foldy-Wal-64,Borie-Rinker-82,Giai-Auerb-81,Con-Don-72,DonPe,Auer-Zam-82,Auer-Kle-84,Urin-Vyaz-92}, essentially only the total muon-capture rates have been measured for a great number of nuclear isotopes \cite{Kol-Lang-94,Kol-Lang-00,Eram-Kuz-Tet-98,Zin-Lang-06,Kuz-Tet-02}.
 On the theoretical side, various nuclear methods using several 
residual interactions allowed the calculation of total capture rates on many nuclei 
with an accuracy of about 10\% compared to the experimental rates. However, for only few isotopes exclusive capture rates to specific states in the daughter nucleus have been determined \cite{Kol-Lang-94,Kol-Lang-00,Eram-Kuz-Tet-98,Kuz-Tet-02}.
As the experimental data for muon capture rates are quite precise, and the theoretical techniques of 
evaluating the nuclear response in the relevant nuclear systems are well developed \cite{Kol-Lang-94,Has4,Ts-Kos-11}, it is worthwhile 
to see to what extent the exclusive capture rates are theoretically understood. 

Furthermore, we mention that, there appear recently, clear indications that the axial-vector coupling constant
$g_{A}=­1.262$ in a nuclear medium is reduced from its free nucleon value \cite{Kol-Lang-94,Kol-Lang-00,Zin-Lang-06,Hau-91,Wild-84,Suh-Civi-13,Ejiri-Suhonen-15}.
 The evidences for such a renormalization of the value $g_A$ come primarily from the analysis of beta decay modes between 
low-lying states of medium-heavy nuclei \cite{Wild-84}
but the use of a quenched $g_{A}$ value is mainly invoked from
the second-order core polarization caused by the tensor force \cite{Bert-Ham-82} and
the screening of the Gamow-Teller (GT) operator by the $\Delta$-hole pairs \cite{Eri-Fig-73}.
Thus, it is necessary to scrutinize on the in-medium quenching of the axial vector 
coupling constant which is in agreement with various well-known indications that $g_A$ is reduced to the value of $g_A \approx 1.000$.
In this work we are not going to study systematically this effect, but we will compare our results of $\mu$-capture rates obtained with the values (i) $g_{A} = 1.262$ and (ii) $g_{A} = 1.135$ with other theoretical ones obtained with the latter value.

The rest of the paper, is organized as follows. 
In Section 2, we summarize briefly, the main characteristics of the effective charged-current weak interaction Hamiltonian and present the main formalism of the ordinary muon capture rates which is based on our compact formalism for the relevant nuclear transition matrix elements (relying on the Donnelly-Walecka projection method) and in the expressions for exclusive, partial and total muon capture rates \cite{DonPe,Don-Wal-76,Has4}. Special focus is given on the calculation of the nuclear wave functions derived within the context of the pn-QRPA. In Section 3, we concentrate on the determination of the required model parameters for the nuclear ground state, derived by solving the BCS (Bardeen Cooper Schrieffer) equations, as well as of the excited states (solution of the pn-QRPA equations).
Our results (Sestion 4) refer to exclusive, partial and total muon capture rates, of the above mentioned nuclear isotopes, which cover the light- and medium-weight region of the periodic table. We also include the individual contributions of the   polar-vector and axial-vector operators in each of the multipole states and in the total Ordinary Muon Carture (OMC) rate.
Finally, in Section 5, we summarize the main conclusions extracted from the present work.

\section{Formalism of muon capture rates}
\label{Formalism of muon capture rates}

The ordinary muon-capture process, that takes place in muonic atoms and  is represented by the semi-leptonic reaction (\ref{mucap}),
proceeds via a charged-current  weak-interaction Hamiltonian which is written as a product of a leptonic, $j_{\mu}^{lept}$, and a hadronic current, $\hat{\mathcal{J}}^{\mu}$, as \cite{DonPe,Don-Wal-76,Has4,Gian-Kos-13}
\begin{eqnarray}
\hat{\mathcal{H}}_{w}=
\frac{G}{\sqrt{2}}j_{\mu}^{lept}\hat{\mathcal{J}}^{\mu}
\end{eqnarray}
where $G=G_{F}cos\theta_{c}$ with $G_{F}$ and $\theta_{c}$ being the well known weak interaction coupling constant and the Cabbibo angle, respectively.

From the nuclear theory point of view, the main task is to calculate the partial and total capture rates of the reaction (\ref{mucap}) which are based on the evaluation of exclusive nuclear transition matrix elements of the form
\begin{eqnarray}
\label{nucl-tran-ME}
\langle f|\widehat{H_{w}}|i\rangle = \frac{G}{\sqrt{2}} \, \ell^{\mu} \int d^{3}x \, e^{-i\textbf{q}\textbf{x}} \langle f|\widehat{\mathcal{J}_{\mu}}|i\rangle.
\end{eqnarray}
(the integration is performed in the region of the nuclear system).
In the latter expression  $| i \rangle $ and $| f \rangle $ denote the initial (ground) and the final nuclear states, respectively. The quantity $\ell^{\mu}e^{-i\textbf{q}\textbf{x}}$ stands for the leptonic matrix element written in coordinate space with $\textbf{q}$ being the 3-momentum transfer. The  
magnitude of ${\bf{\overrightarrow{q}}}$ is defined from the kinematics of the process and is approximately given by \cite{Mukh-Chiang-98}
\begin{eqnarray}
\label{3-momentum transfer}
q \equiv q_{f}= m_{\mu} - \epsilon_{b} + E_{i} - E_{f}
\end{eqnarray}
where $m_{\mu}$ is the muon rest mass, $\epsilon_{b}$ is the muon-binding energy in the muonic atom, $E_{i}$ denotes the energy of the initial state of the parent nucleus and $E_{f}$ the final energy of the corresponding daughter nucleus.

In the unified description of all semi-leptonic electro-weak processes in nuclei developed by  Donnelly and Walecka \cite{Don-Wal-72,Don-Wal-73,Con-Don-72,Don-Wal-76,DonPe}, the calculation of the required transition strengths of Eq. (\ref{nucl-tran-ME}) is based on a multipole decomposition of the hadronic current density which leads to a set of eight independent irreducible tensor multipole operators (four of them come from the polar-vector component and the other four from the axial-vector component of the nuclear current). 
In the present work we assume that the pn-QRPA excitations $|J^{\pi}_{m}\rangle$ have good quantum numbers of angular momentum (J), parity ($\pi$) and energy which is a basic assumption for the Donnelly-Walecka projection method to be applicable.
In this spirit, the computation of each partial transition rate of the muon capture is written in terms of the eight different nuclear matrix elements (between the initial $|J_{i}\rangle $ and the final $|J_{f}\rangle $ states) as
\begin{eqnarray}\label{muon-cap-rates}
\Lambda_{i \rightarrow f} = \frac{2G^{2} q_{f}^{2}}{2J_{i} + 1} R_{f} &\Big[& \big|\langle J_{f} \Vert \Phi _{1s}(\widehat{\mathcal{M}}_{J}-\widehat{\mathcal{L}}_{J})\Vert J_{i}\rangle \big|^{2}\\
& + & \big|\langle J_{f} \Vert \Phi_{1s}(\widehat{\mathcal{T}}_{J}^{el}-\widehat{\mathcal{T}}_{J}^{magn})\Vert J_{i}\rangle \big|^{2} \Big]\nonumber
 \end{eqnarray}
where $\Phi_{1s}$ represents the muon wave function in the 1s muonic orbit \cite{Primak-59}. The operators in Eq. (\ref{muon-cap-rates}) refer to as Coulomb $\widehat{\mathcal{M}}_{J}$, longitudinal $\widehat{\mathcal{L}}_{J}$, transverse electric $\widehat{\mathcal{T}}^{el}_{J}$ and transverse magnetic  $\widehat{\mathcal{T}}^{magn}_{J}$ multipole operators and contain polar-vector and axial-vector parts (see Appendix \ref{Nuclear Matrix Elements}). The factor $R_{f}$ in Eq. (\ref{muon-cap-rates}) takes into consideration the nuclear recoil which is written as 
$R_{f} = \Big( 1 + q_{f}/{M_{targ}}\Big)^{-1}$, 
with $M_{targ}$ being the mass of the target nucleus.
 
\section{Description of the nuclear method}\label{Nuclear Method}

For reliable predictions of partial muon-capture rates,  a consistent description of 
the structure of the ground state $|J_{i}\rangle$ of the parent nucleus as well as of the multipole excitations $|J_{f}\rangle$ of the daughter nucleus are required. 
In the present work, the state-by-state muon capture rates are evaluated using Eq. (\ref{muon-cap-rates}) with the transition 
matrix elements between the states $|J_{i}\rangle$ and $|J_{f}\rangle$  determined with the use of the BCS and pn-QRPA equations, respectively (the BCS 
equations determine the ground state and the pn-QRPA equations provide the excited states as it is shown below) \cite{Has4,Gian-Kos-13,Ts-Kos-11,Bal-Ydr-11,Balasi-Ydr-11,tsak-kos-11,Bal-Ydr-12,kos-ts-12}.
To this end, at first we have chosen the active model space (the same for proton and neutron configurations) for each studied isotope consisted of the single particle j-shells shown in Table \ref{model space}.
\begin{table}[ht]
\caption{The used active model space with the  respective harmonic oscilator parameter for all the studied nuclei. In the last column the major harmonic oscillator shells N plus the individual orbits used for each nucleus are listed.}
\label{model space}
\begin{tabularx}{0.48\textwidth}{c|c|c|c|l}
\hline\hline
& &\multicolumn{3}{c}{Model Space} \\[0.5ex]
\cline{3-5}
Nucleus & b(h.o) &  Core & \begin{tabular}[c]{@{}l@{}}Active\\ Levels\end{tabular} & N ($\hbar\omega$)\\
\hline
  $^{28}Si$  & 1.809 & No       & 10  & 0,1,2,3\\
  $^{32}S$   & 1.843 & No       & 12  & 0,1,2,3,$0g_{9/2}$,$0g_{7/2}$ \\
  $^{48}Ti$  & 1.952 & No       & 12  & 0,1,2,3,$0g_{9/2}$,$0g_{7/2}$ \\
  $^{56}Fe$  & 1.996 & $^{16}O$ & 12  & 2,3,4\\
  $^{66}Zn$  & 2.043 & $^{16}O$ & 12  & 2,3,4\\
  $^{90}Zr$  & 2.138 & $^{16}O$ & 16  & 2,3,4,$0h_{11/2}$,$0h_{9/2}$,$1f_{7/2}$,$1f_{5/2}$\\
\hline\hline
\end{tabularx}
\end{table}


As it is well known, in a rather good approximation, the nucleus can be considered as a system of $Z$ protons and $N$ neutrons
moving independently inside the nuclear volume and attracted by the nuclear center through 
a central strong nuclear force. This central attraction is well described by a mean field which, in our case, is assumed to be 
a Woods-Saxon potential with a Coulomb correction and a spin-orbit parts \cite{Ts-Kos-11}.
%
%
For the latter potential we tested two different parametrizations: i) that of Bohr and Motelson \cite{Bohr-Mot},
and ii) that of the IOWA group \cite{IOWA} and found that both give rather similar results. For the purposes of the present work, however, we adopted the more realistic IOWA parametrization \cite{IOWA}.

For a reliable nuclear Hamiltonian, in addition to the mean field, the two-nucleon correlations, known as residual two-body 
interaction, are necessary to be included. Towards this aim, we employed the
pn-Bonn C-D one-boson exchange potential, but, since the initially evaluated bare
nucleon-nucleon matrix-elements of the latter potential refer to all nuclides with mass number A, for a specific isotope
$(A,Z)$ studied, a renormalization of these two-body matrix elements was carried out with the use of four multiplicative
parameters: The first two, known as pairing parameters $g_{pair}^{p,n}$, for
protons (p) and neutrons (n), renormalize the monopole (pairing) interaction which is the part of the correlations involved at the BCS level for the description of the considered independent quasi-particles.
The third, $g_{pp}$, tunes the particle-particle channel and the
fourth, $g_{ph}$, renormalizes the particle-hole interaction of the Bonn C-D potential. We briefly summarize the adjustement of these parameters below (subsection \ref {pn-QRPA}).

\subsection{Determination of the parent nucleus ground state}
\label{BCS-sec}

The ground state of the parent nucleus, is obtained within the
context of the BCS theory where the one-quasi-particle states are deduced by solving (iteratively) the BCS equations. Towards this aim one is defining quasi-particle creation, $\alpha^{\dagger}$, and annihilation, $\alpha$, operators related to the particle-creation,  $c_{\kappa}^{\dagger}$, and particle annihilation, $c_{\kappa}$, operators through the Bogolyubov-Valatin transformations \cite{Ring-Schuck,Rowe}
\begin{eqnarray}\label{Bog-Val}
\alpha^{\dagger}_{\kappa} = u_{k}c^{\dagger}_{\kappa} - \upsilon_{k} \tilde{c}_{\kappa},\  \tilde{\alpha}_{\kappa} = u_{k} \tilde{c}_{\kappa} + \upsilon_{k}c^{\dagger}_{\kappa},
\end{eqnarray}
where  $\tilde{c}_{\kappa}$ denotes the time reversed particle annihilation operator defined as $\tilde{c}_{\kappa} = (-1)^{j_{k}+m_{k}}c_{-\kappa}$ with $-\kappa=(k,-m_{k})$.
The probability
amplitudes $v_{k}$ and $u_{k}$ for the $k$ single particle level to be occupied
or unoccupied, respectively, are \cite{Ring-Schuck}
\begin{eqnarray}\label{U-V-ampl}
\upsilon^{2 (p,n)}_{k}=\frac{1}{2}\bigg[1-\frac{\epsilon_{k}^{p(n)}-\lambda_{p(n)}}{E_{k}^{p(n)}}\bigg],
\end{eqnarray}
($ u^{2}_{k} = 1 -\upsilon^{2}_{k}$) where $\epsilon_{k}$ is the single particle energy of the $j_{\kappa}$-level and $\lambda_{p}$ ($\lambda_{n}$) denotes the chemical potential for protons (neutrons).
Moreover, the solution of the relevant BCS equations gives the single quasi-particle energies \cite{Ring-Schuck,Kam-Faes-91}
\begin{eqnarray}
E_{k}^{p(n)} = \sqrt{(\epsilon_{k}^{p(n)}-\lambda_{p(n)})^{2} +\Delta_{k}^{2}}
\end{eqnarray}
with $\Delta_{\kappa}$ being the theoretical energy gaps  
$(\Delta^{k}=-\sum_{k^{\prime}>0}\bar{\upsilon}_{k\bar{k}{{k^{\prime}\bar{k}^{\prime}}}}
u_{k^{\prime}}\upsilon_{k^{\prime}})$ 
\cite{Ring-Schuck}.
From the solution of the gap equation \cite{Kam-Faes-91,Bar-60}
\begin{eqnarray}
\label{theoretical gaps}
\Delta^{k}_{p(n)} = \frac{g_{pair}^{p(n)}}{2[j_{k}]}\sum_{k^{\prime}} [j_{k^{\prime}}] \frac{\Delta_{k^{\prime}}}{ E_{k}^{p(n)}} \langle (kk)0|\mathcal{G}|(k^{\prime}k^{\prime})0\rangle 
\end{eqnarray}
(here the notation is, $[j] = \sqrt{2j+1}$) one obtains the pairing gaps for protons $\Delta^{k}_{p}$ and neutrons $\Delta^{k}_{n}$
through the renormalization of the proton and neutron pairing matrix elements $ \langle (kk)0|\mathcal{G}|(k^{\prime}k^{\prime})0\rangle $ of the residual interaction, using the parameters $g_{pair}^{p}$ and $g_{pair}^{n}$. The lowest quasi-particle energy, obtained from the gap equation, is 
determined, through the pairing parameters $g_{pair}^{p(n)}$ entering the theoretical gaps of Eq. (\ref{theoretical gaps}) so as to reproduce the experimental (empirical) energy gaps $\Delta_{p,n}^{exp}$ given from the three point formula \cite{Kam-Faes-91}
\begin{eqnarray}
\Delta^{exp}_{p (n)} =-\frac{1}{4}&\Big[&S_{p (n)}[(A-1,Z-1 (Z))]-2S_{p (n)}[(A,Z)]\nonumber\\
&+& S_{p (n)}[(A+1,Z+1 (Z))]\Big].
\end{eqnarray}
In the latter equation $S_{p}$ and $S_{n}$ are the experimental separation energies for protons and neutrons, respectively, of the target nucleus (A,Z) and of the neighboring nuclei $(A\pm 1, Z\pm 1)$ and $(A\pm 1, Z)$. 
Here, we used the method of Ref. \cite{Kam-Faes-91} to obtain 
the $g_{pair}^{p,n}$ values for the studied nuclei and tabulate them in Table \ref{BCS}.
We note that, in order to achieve the reproducibility of the experimental energy spectrum  in similar QRPA calculations some authors modify slightly the Woods-Saxon proton and neutron single particle energies in the vicinity of the nuclear Fermi surfaces \cite{Dean-Lang-98,Bal-Ydr-11}. In this work, we pay special attention on the reproducibility of the energy spectrum of the daughter nucleus as is discussed in detail in the next section.

\begin{table}[h]
\caption{\label{BCS}Parameters for the renormalization of the interaction of proton
pairs, $g_{pair}^{p}$, and neutron pairs, $g_{pair}^{n}$. They have been
fixed in such a way that the corresponding experimental gaps,
$\Delta_{p}^{exp}$ and $\Delta_{n}^{exp}$, are quite accurately reproduced.}
\begin{small}
\begin{center}
\begin{tabular}{lllllll}
\hline \hline\\[-0.3cm]
Nucleus &$g_{pair}^{n}$& $g_{pair}^{p}$  &\begin{tabular}[c]{@{}l@{}}$\Delta_{n}^{exp}$\\ (MeV) \end{tabular}  & \begin{tabular}[c]{@{}l@{}}$\Delta_{n}^{theor}$\\(MeV)\end{tabular}  &\begin{tabular}[c]{@{}l@{}} $\Delta_{p}^{exp}$\\ (MeV)\end{tabular} & \begin{tabular}[c]{@{}l@{}} $ \Delta_{p}^{theor}$\\ (MeV)\end{tabular} \\[0.5ex]
\hline\\[-0.3cm]
    $^{28}Si$   & 1.1312 & 1.0601  & 3.1428 & 3.1429 & 3.0375 & 3.0377\\[0.5ex]
    $^{32}S$    & 0.8862 & 0.8230  & 2.0978 & 2.0979 & 2.0387 & 2.0386\\[0.5ex]
    $^{48}Ti$   & 0.9259 & 0.9833  & 1.5576 & 1.5578 & 1.9112 & 1.9111 \\[0.5ex]
    $^{56}Fe$   & 0.9866 & 0,9756  & 1.3626 & 1.3626 & 1.5682 & 1.5683 \\[0.5ex]
    $^{66}Zn$   & 1.0059 & 0.9271  & 1.7715 & 1.7716 & 1.2815 & 1.2814 \\[0.5ex]
    $^{90}Zr$   & 0.9057 & 0.7838  & 1.8567 & 1.8568 & 1.1184 & 1.1183 \\[0.5ex]
\hline \hline
\end{tabular}
\end{center}
\end{small}
\end{table}
\subsection{The pn-QRPA excitation spectrum of the daughter nucleus}
\label{pn-QRPA}
For the purposes of the present study, transitions between the $|0^{+}\rangle$ ground state of a rather spherical even-even parent-nucleus and the excited 
states of the resulting daughter nucleus are the basic ingredients.
For several charged-current reactions, the pn-QRPA method provides a reliable description of the nuclear excited states of the resulting odd-odd nuclear system in Eq. (\ref{mucap})
 \cite{Kam-Faes-91}. Here, we exploit this advantage in order to derive the excitation spectrum of the daughter nucleus produced in the $\mu$-capture process. In this context, we first define the two quasi-fermion operators $A^{\dagger}$ and $\tilde{A}$ (which obey boson commutation relations in a correlated RPA ground state) as \cite{Ring-Schuck,Rowe,Kosmas-01,Kos-Verd-94,Kos-Faes-97,Ts-Kos-11,Bal-Ydr-11,Balasi-Ydr-11,tsak-kos-11,Bal-Ydr-12,kos-ts-12,Has4,Gian-Kos-13}
\begin{eqnarray}
A^{\dagger}_{mi} (JM) &=& [a_{j_{m}}^{\dagger}a_{j_{i}}^{\dagger}]^{J}_{M}\\
 &=& \sum_{m_{m}(m_{i})} \langle j_{m}j_{i}m_{m}m_{i}|JM\rangle \alpha^{\dagger}_{j_{m}m_{m}} \alpha^{\dagger}_{j_{i}m_{i}},\nonumber
\end{eqnarray}

\begin{eqnarray}
\tilde{A}_{mi}(JM) &=& (-1)^{J-M}A_{mi}(J-M).
\end{eqnarray}
 Afterwards, we write down the pn-QRPA phonon operators
\begin{eqnarray}
\label{phon-oper}
Q^{\nu\dagger}_{J^{\pi}M} =\sum _{m\leq i} [X^{\nu}_{mi}A^{\dagger}_{mi}(JM) + Y^{\nu}_{mi}\tilde{A}_{mi}(JM)],
\end{eqnarray}
$\nu$ enumerates the multipole states of the multipolarity $J^{\pi}$,
that creates the excitation $|\nu\rangle \equiv|J^{\pi}_{\nu}\rangle$ by acting on the QRPA vacuum $|\tilde{0}\rangle_{QRPA}$ as \cite{Kosmas-01,Kos-Faes-97,Ts-Kos-11,Bal-Ydr-11,Balasi-Ydr-11,tsak-kos-11,Bal-Ydr-12,kos-ts-12}
\begin{eqnarray}
|J^{\pi}_{\nu}\rangle = Q^{\nu\dagger}_{J^{\pi}M} |\tilde{0}\rangle_{QRPA}.
\end{eqnarray}

The X (forward) and Y (backward) scattering amplitudes entering Eq. (\ref{phon-oper}) are obtained by solving the pn-QRPA equations (pn-QRPA eigenvalue problem) which in matrix form is written as \cite{Ring-Schuck}
\begin{eqnarray}\label{matr-form-QRPA}
\left(\begin{array}{cc}
        \mathcal{A} & \mathcal{B}\\
        \mathcal{-B} & \mathcal{-A}
          \end{array}\right)
\left(\begin{array}{c}
        X^{\nu} \\
        Y^{\nu} 
          \end{array}\right)=\Omega^{\nu}_{J^{\pi}}\left(\begin{array}{c}
        X^{\nu} \\
        Y^{\nu} 
          \end{array}\right),
\end{eqnarray}
$\Omega^{\nu}_{J^{\pi}}$ denotes the excitation energy of the QRPA state $|J^{\pi}_{\nu}\rangle$. Thus, the X and Y amplitudes are calculated seperately for each multipole set of states (multipolarity).

The reliability of the QRPA excitations $\Omega^{\nu}_{J^{\pi}}$ and of the corresponding  many-body nuclear wave functions is checked through the reproducibility of the energy spectrum of the final odd-odd nucleus.
The values of particle-particle $(g_{pp})$ and particle-hole $(g_{ph})$ parameters in the set of isotopes chosen (determined separately for each multipolarity) \cite{Ts-Kos-11,tsak-kos-11,kos-ts-12} lie in the region 0.65 - 1.20 (with the exception of the $1^+$ and $2^-$ multipolarities in some isotopes, for which the values are rather small, 0.2 - 0.6)  \cite{PGian-Kos-13}.
Such small values of the strength parameters come out  in studies of charged current reactions ($e^-$-capture, single- and double-beta decays) when fitting simultaneously the QRPA parameter, $g_{pp}$, and the axial vector coupling constant, $g_A$ \cite{Sim-Faes-08,Faes-Fogli-08,Rod-Faes-06,Suh-Civi-13,Faess-09}.
We stress that in our QRPA method the strength parameters are determined through the reproduction of the energy spectrum of the daughter nucleus but we have also made an effort to test them through the GT energy position and the total GT strength \cite{Kuz-Tet-02,Lang-Pin-03}. Even though our GT-type operator contributes differently (due to the presence of the Bessel function), we found that, the total GT strength differs significantly (more than a factor of 2.5) from the experimental one, although, the energy position is well reproduced. 
In our muon capture (and $e^-$-capture) rates the simultaneous  variation of $g_A$ and $g_{pp}$ parameters has not been checked extensively (see Ref. \cite{Gian-Kosm-15}.

\begin{figure*}[!t]
\begin{center}
\includegraphics[scale = 1.0]{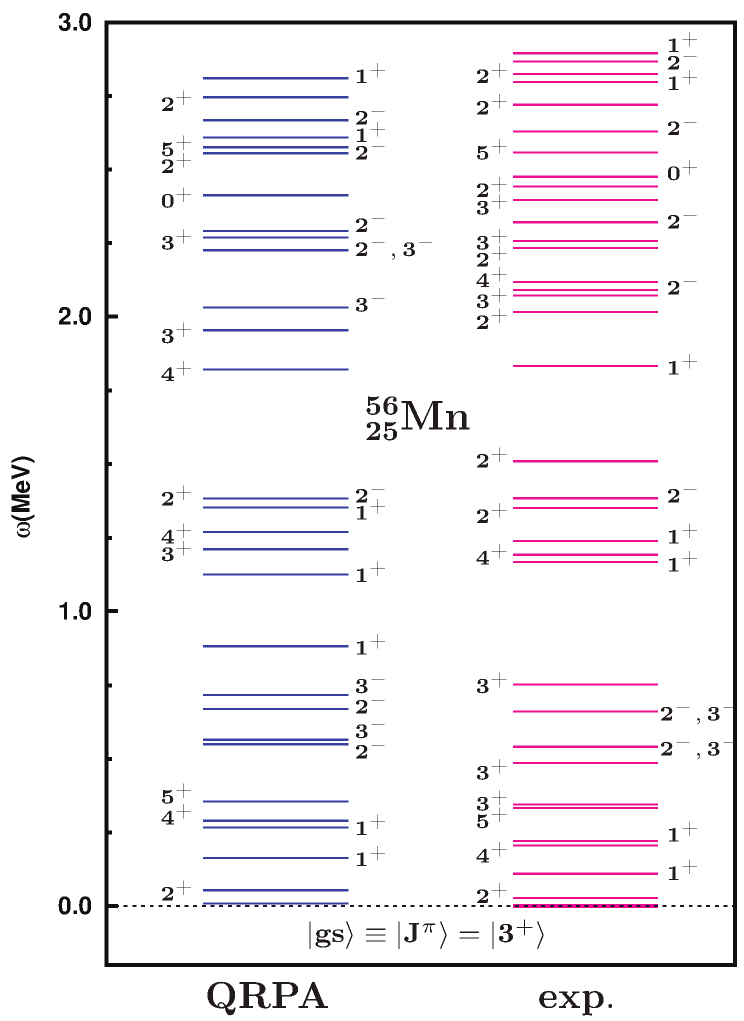}
\includegraphics[scale = 1.0]{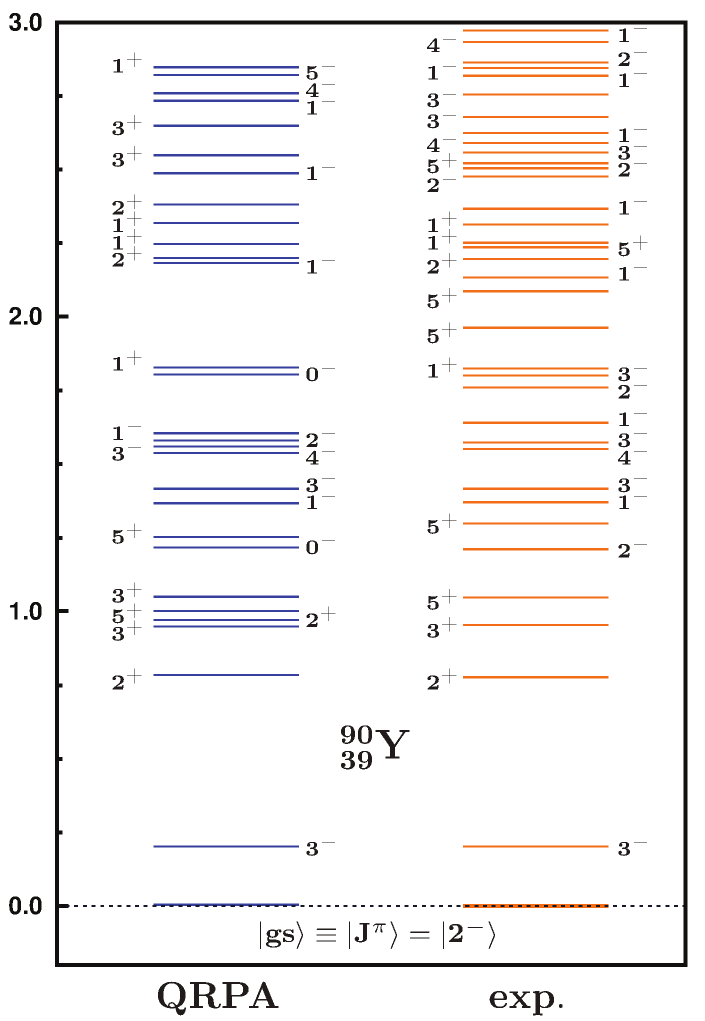}
\end{center}
\caption{Comparison of the theoretical excitation spectrum, resulting from the solution of the pn-QRPA eigenvalue problem, with the low-lying (up
to about 3 MeV) experimental one for $^{56}Mn$ and $^{90}Y$ nuclei (for the other spectra see Ref. \cite{Giannaka,PGian-Kos-13}). The agreement is quite good at least for low excitation energies.}
\label{fig-Fasma}
\end{figure*}

We furthermore note that, in order to achieve the reproducibility of the experimental energy spectrum of the daughter nucleus and for  measuring the excitation energies of the daughter nucleus from the ground state of the initial (even-even) nucleus, some authors shift the entire set of QRPA spectrum by about $\lambda_{p} - \lambda_{n}$ in the muon capture process \cite{Eram-Kuz-Tet-98}.
In our present study we also adopt the latter treatment, so, the calculated pn-QRPA energy spectrum of each individual multipolarity $J^{\pi}$ is shifted in such a way that the first calculated value of each multipole state (i.e. $1_{1}^{+}, 2_{1}^{+}$...etc), to approach as close as possible to the corresponding lowest experimental energy of the daughter nucleus. Such a shifting is necessary whenever in the pn-QRPA a BCS ground state is used, a treatment adopted by other groups too \cite{Yous-Faes-09,Rod-Faes-06}.
Table \ref{shift} shows the shifting applied to the QRPA spectrum for each multipolarity of the studied nuclei.
\begin{table*}
 \caption{The shift of the spectrum seperately of each state in MeV}
 \label{shift}
 \begin{center}
 \begin{tabular}{c|c|c|c|c|c|c||c|c|c|c|c|c|c}
 \hline
 \hline
 \multicolumn{7}{c}{Positive Parity States} & \multicolumn{6}{c}{Negative Parity States} \\[0.5ex]
 \hline\\[-0.3cm]
 $J^{+}$ & $^{28}Si$ & $^{32}S $ & $^{48}Ti $ & $^{56}Fe $ & $^{66}Zn $ & $^{90}Zr $& $J^{-}$  & $^{28}Si$ & $^{32}S $ & $^{48}Ti $ & $^{56}Fe $ & $^{66}Zn $ & $^{90}Zr $ \\[0.5ex]
 \hline
 $0^{+}$ & 2.60 & 0.00 & 0.65 & 1.60 & 0.90 & 1.00 & $0^{-}$ & 4.20 & 1.00 & 4.00 & 4.30 & 5.00 & 4.47 \\[0.5ex]
 $1^{+}$  & 5.00 & 2.50 & 2.65 & 5.90 & 2.50 & 2.85 & $1^{-}$ & 4.40 & 4.05 & 4.00 & 4.20 & 6.80 & 4.30 \\[0.5ex]
 $2^{+}$ & 4.35 & 2.43 & 2.10 & 3.10 & 2.55 & 2.78 & $2^{-}$ & 5.80 & 4.40 & 5.10 & 6.80 & 3.85 & 2.39 \\[0.5ex]
 $3^{+}$ & 5.90 & 0.00 & 2.70 & 2.30 & 2.50 & 2.82 & $3^{-}$ & 6.00 & 3.98 & 4.10 & 6.80 & 2.60 & 2.59 \\[0.5ex]
 $4^{+}$ & 4.90 & 3.56 & 3.25 & 2.50 & 1.75 & 0.00 & $4^{-}$ & 5.00 & 2.57 & 4.25 & 3.50 & 3.55 & 1.30 \\[0.5ex]
 $5^{+}$ & 2.70 & 0.84 & 3.35 & 2.00 & 0.55 & 2.40 & $5^{-}$ & 6.50 & 0.00 & 3.05 & 3.50 & 3.00 & 0.00 \\[0.5ex]
 \hline \hline
 \end{tabular}
  \end{center}
 \end{table*}
We note that, a similar treatment is required in QRPA calculations for double-beta decay studies where the excitations derived for the intermediate odd-odd nucleus (intermediate states) through p-n or n-p reactions from the neighboring nuclei do not match each other \cite{Rod-Faes-06}. 
The resulting low-energy spectrum (up to 3.0 MeV) using our pn-QRPA method,  agrees well with the experimental one as can be seen from Fig. \ref{fig-Fasma}.

Before proceeding to our results, it is worthwhile to briefly summarize the advantages of the calculational procedure followed in performing the present detailed calculations of partial and total muon capture rates as compared to the methods used by  other groups \cite{Foldy-Wal-64,Eram-Kuz-Tet-98,Kol-Lang-94,Zin-Lang-06}.
In the earlier pioneering work of Foldy and Walecka \cite{Foldy-Wal-64}, the authors related the dipole capture rate to the experimental photoabsorption cross section and used symmetry arguments to compare polar-vector and axial-vector matrix elements. 
The afore mentioned authors derived $\mu^-$-capture sum rules based on the GDR strength excited after $\mu^-$-capture. The required GDR amplitudes are obtained (for light and medium nuclei) from the corresponding photo-absorption cross sections. 
Later, on the calculations of Eramzhyan et al.  \cite{Eram-Kuz-Tet-98} employed a truncated model space with ground state correlations and adopted the standard free nucleon coupling constants. 
In the work of Kolbe et al. \cite{Kol-Lang-94}, for the calculations of muon capture rates, use of the continuum RPA method was made with the free nucleon form factors, while
recently, Zinner et al. \cite{Zin-Lang-06} proposed the use of a quenched value for the axial-vector coupling constant $g_{A}$ in order to reliably evaluate the true  Gamow-Teller transitions.

It is worth mentioning that, recent studies of single and double beta-decays as well as of neutrino-nucleus reactions under stellar conditions, have demonstrated an important role of the quenched value of the coupling constant $g_A$ \cite{Suh-Civi-13,Ejiri-Suhonen-15,Giannaka}.
In the present calculations we also use a quenched value of $g_{A}$ same for all multipole transitions (see the following section).

\section{Results and Discussion}
\label{Results and Discussion}

In the ordinary muon capture on complex ($A\geq 12$) nuclei, the nuclear response is governed by
the momentum transfer q of Eq. (\ref{3-momentum transfer}), i.e. by an  energy transfer to the daughter nucleus of the order of the muon mass $m_{\mu}$ minus the binding energy $\epsilon_{b}$
of the muon in the muonic atom restricted
from below by the mass difference of the initial and final nuclei and from above by 
the muon mass [see Eq. (\ref{3-momentum transfer})]. The phase space and the nuclear response favor lower nuclear excitations, namely the nuclear states in the giant resonance region (GDR and GT resonance) are expected to dominate \cite{Kol-Lang-94}.

In our calculational procedure we followed three steps.
(i) In the first step we performed realistic state-by-state calculations on exclusive OMC rates in the isotopes $ ^{28}Si, ^{32}S, ^{48}Ti, ^{56}Fe, ^{66}Zn $ and $^{90}Zr$, a set which covers a rather wide range of the periodic Table from light- to medium-weight nuclei. These calculations have been performed twice: Once with the use of the free nucleon coupling constants $g_{A} = 1.262$ and the other with the use of the value $g_{A} = 1.135$, to take into acount the rather small quenching effect indicated for medium-weight nuclei \cite{Zin-Lang-06,Hau-91,Wild-84}. We also focused on the study of the relative strength of the polar-vector and axial-vector contributions for each individual excitation induced by the respective components of the muon-capture operators.
(ii) In the second step of our calculations, we examined the dominance of the low-spin multipolarities into the total $\mu$-capture rate. We also estimated the percentage (portion) of their contribution in the total rate for the most important multipolarities.
(iii) In the last step, we evaluated total muon-capture rates for the above set of isotopes.
For all the above calculations, the required wave functions (for the initial (ground) state and for all accessible  final states) were constructed by solving the BCS and QRPA equations, respectively, as described before (see Sections \ref{BCS-sec} and \ref {pn-QRPA}).
\subsection{State by State calculations of exclusive transition Rates in $\mu$-capture}

At first, we evaluated the exclusive $\mu^{-}$-capture rates $\Lambda_{i \rightarrow f}$ of Eq (\ref{muon-cap-rates}) for all multipolarities with $J^{\pi}\leq 5^{\pm}$.
In Eq. (\ref{muon-cap-rates}) transitions between the ground state $\vert i \rangle \equiv \vert 0^{+}_{gs}\rangle$ of a spherical target nucleus and an excited state $\vert J_{f}^{\pi} \rangle \equiv \vert f \rangle$ of the resulting odd-odd nucleus are considered.
In most of the previous studies a mean value of the muon wave function, $\Phi^\mu (\overrightarrow{r})$, with $\overrightarrow{r}$ being 
the spherical coordinate, has been utilized (see Appendix \ref{Wave function}). An accurate description of the reaction
(\ref{mucap}) (and of any reaction having the same initial state with it, i.e. a muon orbiting 
around an atomic nucleus $(A,Z)$), however, requires the exact muon wave function derived by 
solving the Schr\"odinger equation (or the Dirac equations) that obeys a bound muon within 
the extended Coulomb field of the nucleus in such muonic atoms \cite{Zin-Lang-06}.
 
Assuming, that the muon wave function in the region of the nuclear target is nearly constant, the integrals entering Eq. (\ref{muon-cap-rates}) can be performed by taking out of them an average value $\langle \Phi_{1s}\rangle$. Hence, the exclusive muon capture rates $\Lambda_{J_{f}^{\pi}}$ can be rewritten as:
\begin{eqnarray}
 \label{OMC-rates}
\Lambda_{gs \rightarrow J_{f}^{\pi}} \equiv  \Lambda_{J_{f}^{\pi}} 
&=& 2G^{2}\langle \Phi _{1s}  \rangle^{2} R_{f} q_{f}^{2}\cdot\nonumber\\ & \Big[ &\big|\langle J_{f} ^{\pi}\Vert (\widehat{\mathcal{M}}_{J}-\widehat{\mathcal{L}}_{J})\Vert 0_{gs}^{+}\rangle \big|^{2}\nonumber\\ 
 & + & \big|\langle J_{f} ^{\pi}\Vert (\widehat{\mathcal{T}}_{J}^{el}-\widehat{\mathcal{T}}_{J}^{magn})\Vert 0_{gs}^{+}\rangle \big|^{2} \Big]
\end{eqnarray}

\begin{figure*}
\begin{center}
\includegraphics[scale = 0.75]{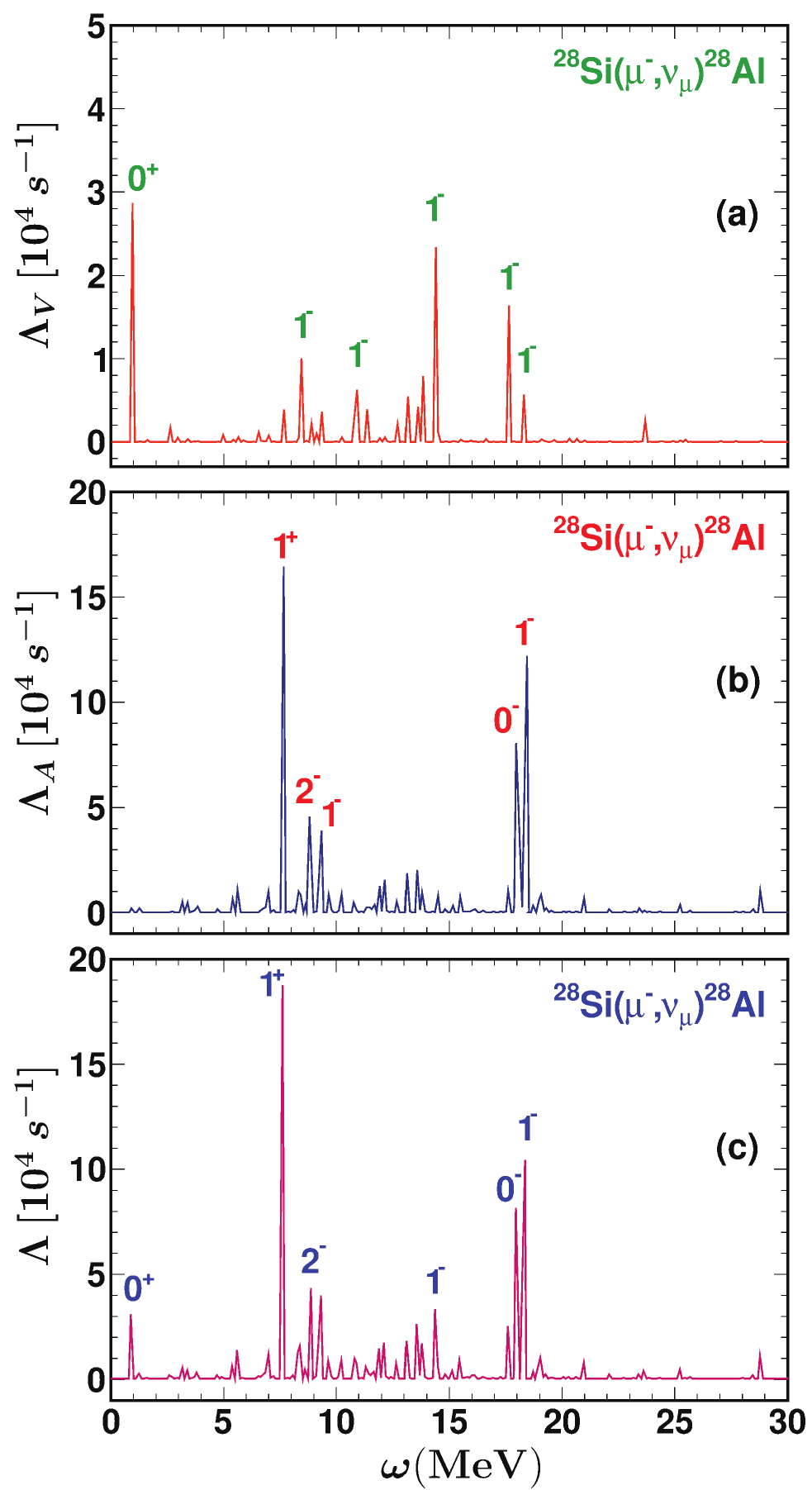}
\includegraphics[scale = 0.75]{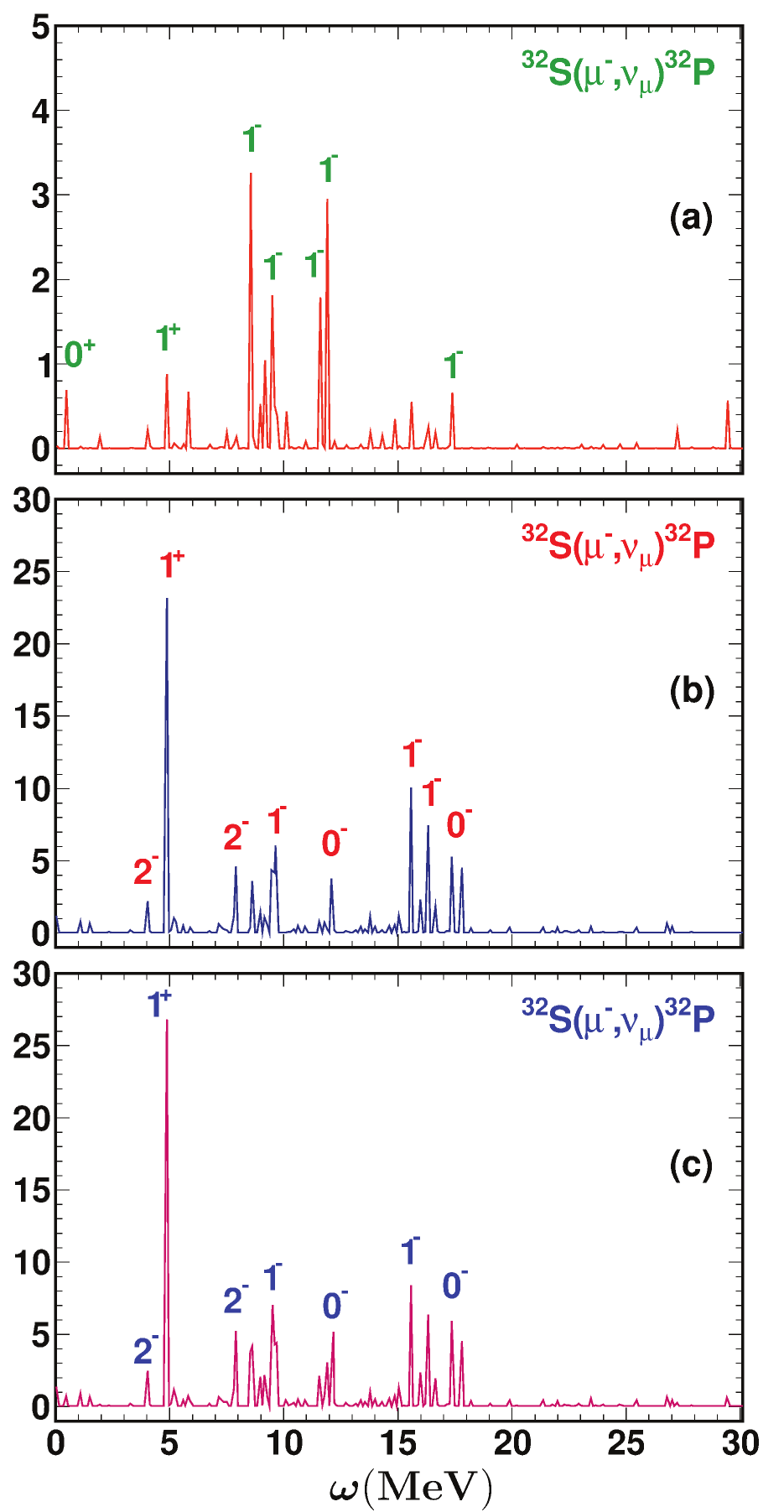}
\end{center}
\caption{Individual contribution of the Polar-Vector $\Lambda_{V}$ (pannel(a)) and Axial-Vector $\Lambda_{A}$(pannel (b)) to the total muon-capture rate (pannel(c)) as a function of the excitation energy $\omega$  for the $^{28}Si$ and $^{32}S$ nuclei.}
\label{Si28-S32-Sort}
\end{figure*}

\begin{figure*}
\begin{center}
\includegraphics[scale = 0.75]{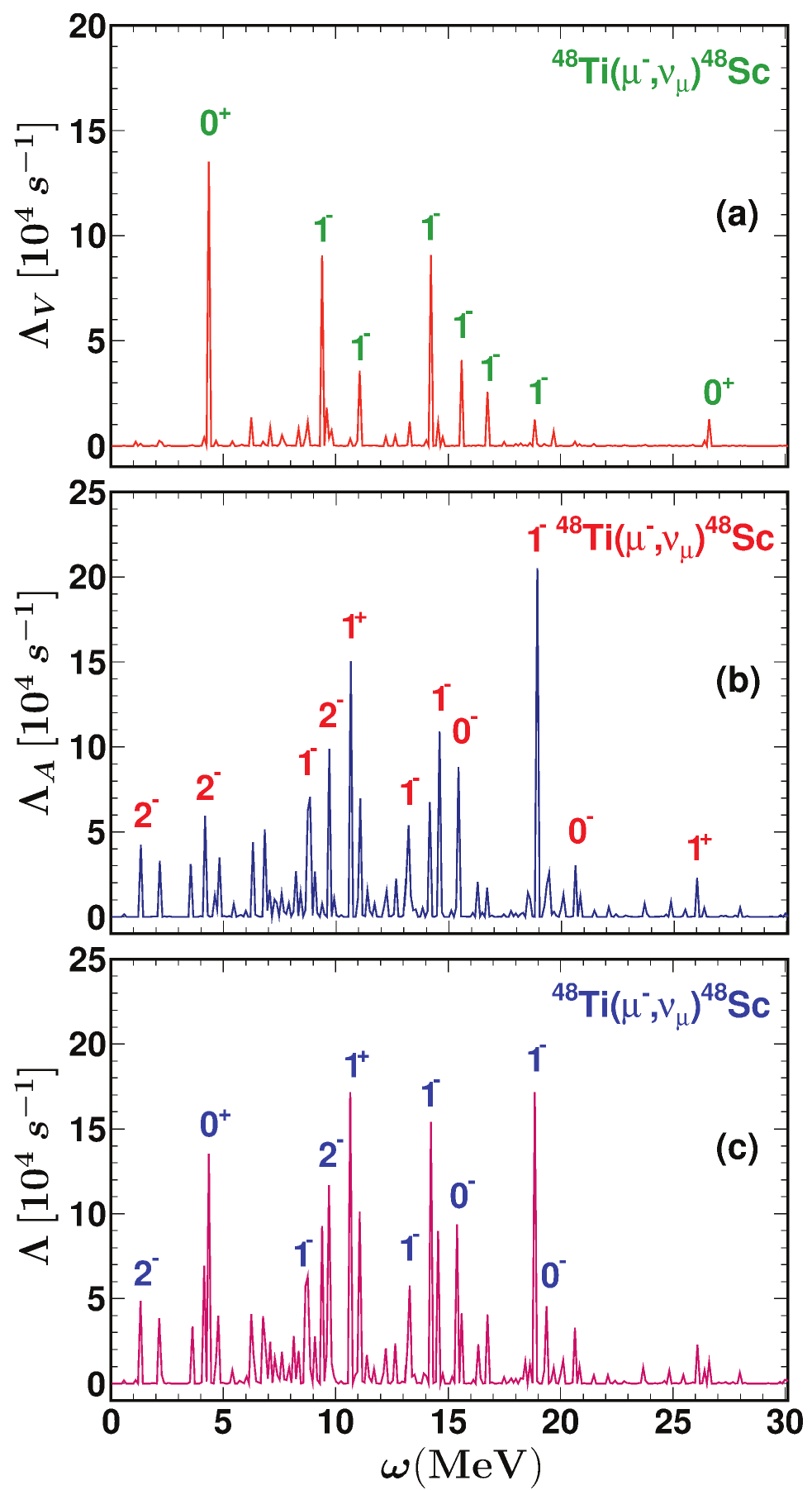}
\includegraphics[scale = 0.75]{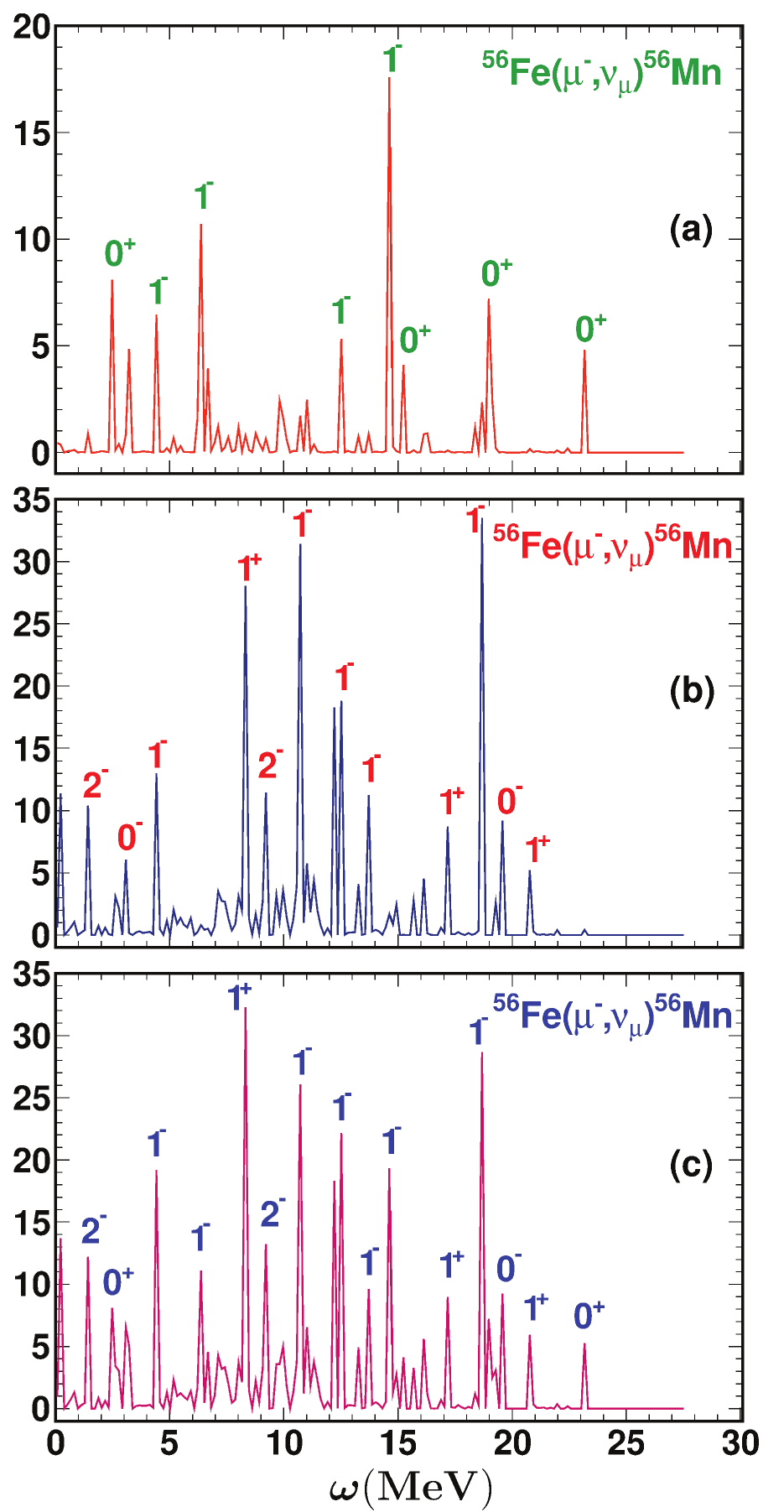}
\end{center}
\caption{The same as Fig. \ref{Si28-S32-Sort} but for the nuclei $^{48}Ti$ and $^{56}Fe$.}
\label{Ti48-Fe56-Sort}
\end{figure*}

\begin{figure*}
\begin{center}
\includegraphics[scale = 0.75]{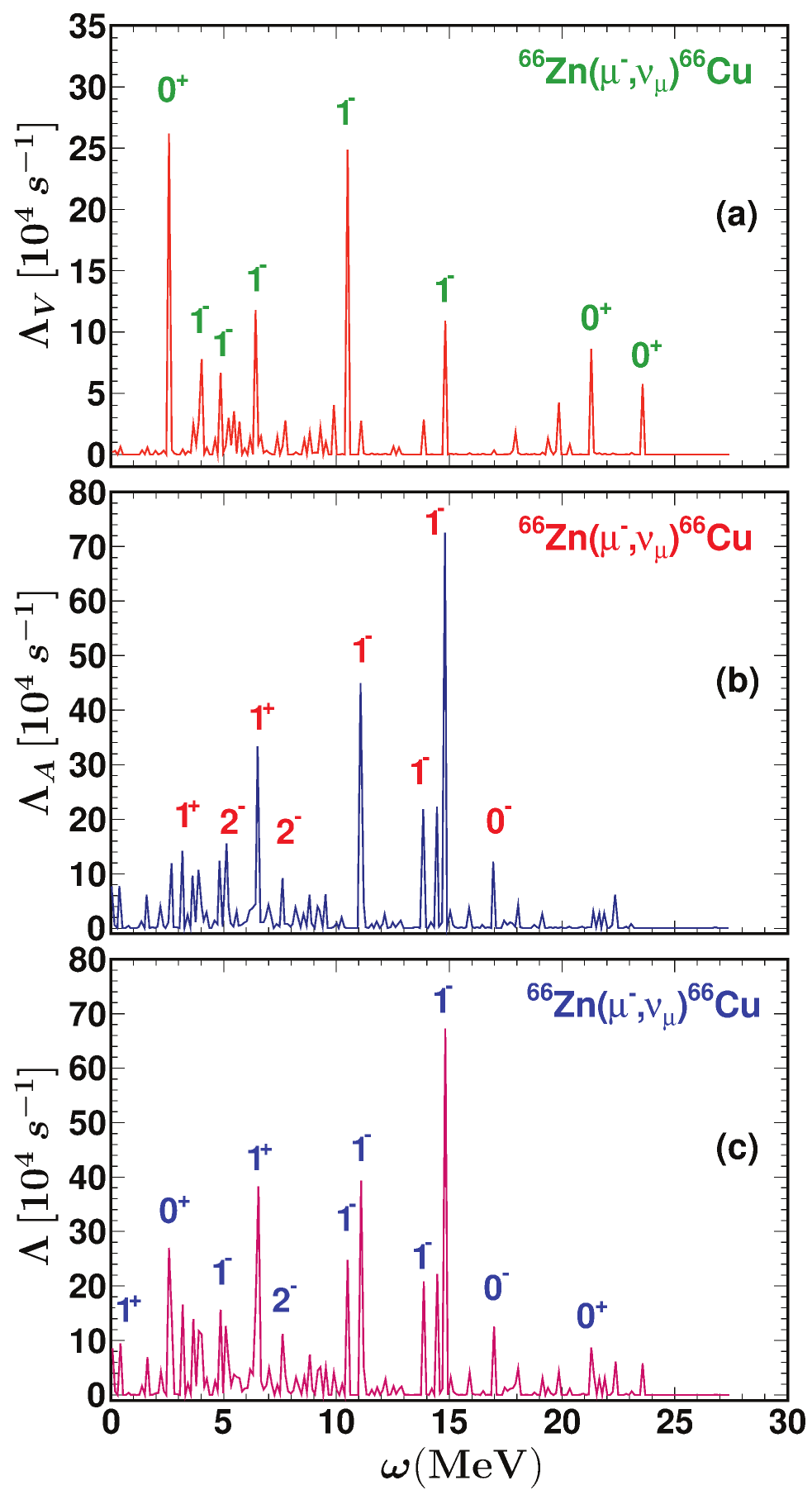}
\includegraphics[scale = 0.75]{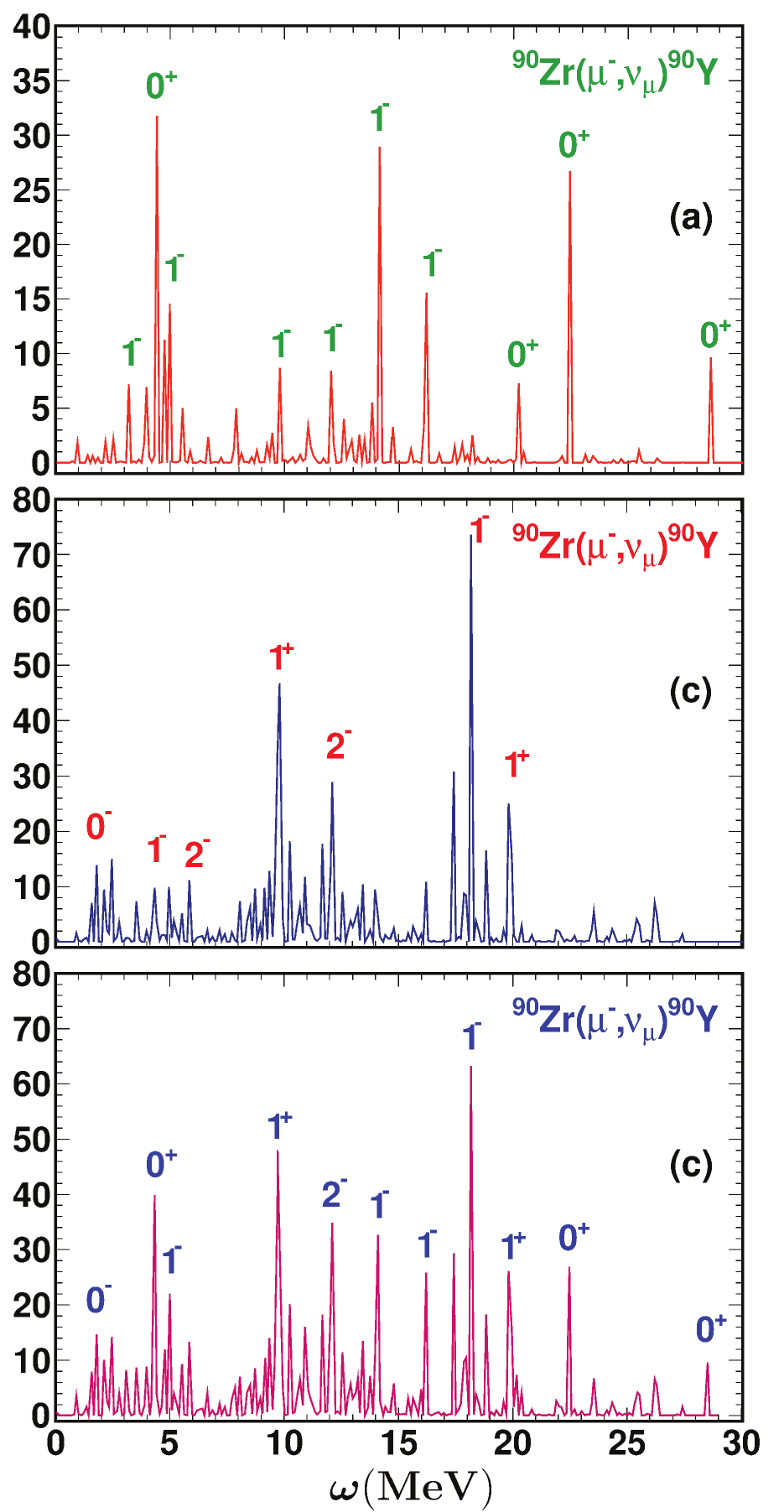}
\end{center}
\caption{The same as Fig. \ref{Si28-S32-Sort} but for the nuclei $^{66}Zn$ and $^{90}Zr$.}
\label{Zn66-Zr90-Sort}
\end{figure*}

On the basis of the latter expression, we initially, performed state-by-state calculations, for the above mentioned set of nuclear isotopes, by using the free nucleon coupling constant $g_{A}$ for the axial-vector form factor.
Then, we repeated these calculations (with the exception of $ ^{28}Si$ and $^{32}S$ isotopes) by taking into account the quenching effect of the axial-vector coupling constant $g_{A} = 1.135$.
For each excitation of the daughter nucleus, our code provides us with the separate contributions induced by the components of the muon-capture operator.
Relying on this possibility, we examined the multipole decomposition of the QRPA response in the muon capture reaction for the studied nuclei. In Figs. \ref{Si28-S32-Sort}, \ref{Ti48-Fe56-Sort} and \ref{Zn66-Zr90-Sort} we illustrate the contribution of each individual transition. We also show the contribution of the polar-vector as well as the axial-vector parts originated from the corresponding components of the weak interaction Hamiltonian (see Sect. \ref{Formalism of muon capture rates}). Evidently, most of the muon capture strength goes to $1^{-}$, $1^{+}$ and $2^{-}$ low-lying multipole excitations, of the particle bound spectrum and of the giant dipole, spin  and spin-dipole resonances. 



As mentioned before, our code initially gives results for exclusive muon capture rates, $\Lambda_{J_{f}^{\pi}}$, seperately for each multipolarity (in ascending order with respect to the pn-QRPA excitation energy $\Omega^{\nu}_{J^{\pi}}$). In order to study the dependence of the rates on the excitation energy $\omega$ throughout the entire pn-QRPA spectrum of the daughter isotopes, a rearrangement of all possible excitations in ascending order with respect to $\omega$ and with the corresponding rates, is required. This was performed by using a special code (appropriate for matrices). Totally, for $J^{\pi}\leq 5^{\pm}$ in the model space chosen for each isotope, we have 286 states for the $^{28}Si$  isotope, 440 states for each of the $^{32}S$ and $^{48}Ti$ isotopes, 488 states for each of the $^{56}Fe$ and $^{66}Zn$,  and 912 states for $^{90}Zr$ isotope in the corresponding daughter nucleus. The variation of the exclusive rates throughout the entire excitation spectrum of the daughter nucleus in the case of the above target isotopes are demonstrated in Figs. \ref{Si28-S32-Sort}, \ref{Ti48-Fe56-Sort} and \ref{Zn66-Zr90-Sort}.
For all reactions, the rates present some characteristic clearly pronounced peaks at various excitation energies $\omega$ and specifically for transitions $J^{\pi} = 1^{+}$, $1^{-}$ but also for $J^{\pi} = 0^{+}$, $0^{-}$ and $2^{-}$ transitions.

More specifically, in the daughter $^{28}Al$ isotope the maximum peak corresponds to the $1^{+}_{7}$ QRPA transition at $\omega = 7.712 \, MeV$ (see Fig. \ref{Si28-S32-Sort}). Other two characteristic peaks are at $\omega = 18.135 \, MeV$ and at $\omega = 18.261 \, MeV$ which correspond to the $0^{-}_{9}$ and $1^{-}_{26}$ transitions respectively.
In the case of $^{32}P$ isotope the maximum peak corresponds to the $1^{+}_{5}$ transition at $\omega = 4.855 \, MeV$. Another characteristic peak is at $\omega = 15.564 \, MeV$ which corresponds to the $ 1^{-}_{28}$ transition as shown in Fig. \ref{Si28-S32-Sort} (left).
For the $^{48}Sc$ isotope, the pronounced peaks correspond to the first excited $0^{+}$ state ($0^{+}_{1}$) (at $\omega = 4.319\, MeV$), the $2^{-}_{17}$ (at $\omega = 9.672\, MeV$), the $1^{+}_{13}$  ($\omega = 10.666\, MeV$) and the $1^{-}_{26}$ transitions ($\omega = 18.868 \,MeV$).
From Fig. \ref{Ti48-Fe56-Sort} (right pannel), for the daughter isotope $^{56}Mn$, we see that the maximum peak appears at $\omega = 8.278\, MeV$ and corresponds to the $1^{+}_{10}$ transition. Another important transition is that of $1^{-}_{38}$ at $\omega = 18.716 \, MeV$.
As shown in Fig \ref{Zn66-Zr90-Sort}, in the case of the daughter isotope $^{66}Cu$, the maximum peak appears at $\omega = 6.555 \, MeV$ and corresponds to  $1^{+}_{10}$ state and a pronounced peak for the $1^{-}_{38}$ at $\omega = 14.833 \, MeV$.
Finally, for the $^{90}Y$ isotope, the maximum peak appears for the $1^{-}_{54}$ transition  at $\omega = 18.218\, MeV$ and for the $1^{+}_{36}$ at $\omega = 9.752 \,MeV$.

From the above results, we conclude that in general, a great part of the OMC rate comes from the excitation energy region where the centroid of the GT strength is located for each daughter nucleus. As it is known from closure approximation studies \cite{Rose-77,Rose-78}, the mean excitation energy in muon capture (about 15 MeV) is nearly equal to the energy of the giant dipole resonance (GDR) which is slowly decreasing with A or Z \cite{Zin-Lang-06}. On the other hand, the GT-like operators (in which the full spherical Bessel functions is taken into account) contribute very little in heavier nuclei where most of the active neutrons and protons are in different oscillator shells. In lighter nuclei, however, i.e. for nuclei having N and Z smaller than 40, the GT strength is significant and it is concentrated at the low energy region. Regarding the giant spin resonance ($J^{\pi}=1^{+}$) for all nuclei the peak of the exclusive $\mu$-capture rate is located between 5-11 MeV.
It should be stressed that concerning the pronounced contribution to the $1^{-}$ states, it may contain a small portion of the spurious center of mass motion part (up to about $17\%$ in our QRPA method) \cite{Ts-Kos-11}. 
This is due to the isoscalar movement of the nucleons in the mean field (dipole oscillation of the whole nucleus). 
As it is known this is usually removed by using specific methods \cite{Ts-Kos-11}.

As it becomes clear from  Figs. \ref{Si28-S32-Sort}, \ref{Ti48-Fe56-Sort} and \ref{Zn66-Zr90-Sort}, for the studied nuclei the muon capture response presents maximum peak in the very important giant dipole resonance region (GDR), which is located in the energy region of 18-19 MeV for $^{28}Si$, $^{48}Ti$, $^{56}Fe$ and $^{90}Zr$ isotopes, and in the region of 15-16 MeV for $^{32}S$ and $^{66}Zn$ isotopes.
These results can be compared with the empirical expression, for medium-weight and heavy isotopes, which gives the energy location of the giant dipole resonance, $E_{IVD}$, based on the Jensen-Steinwedel and Goldhaber-Teller models (a hydrodynamical view of the giant resonance) as \cite{Har-01}
\begin{eqnarray}
\label{giant-dipole-resonance}
E_{IVD} = 31.2 A^{-1/3} + 20.6 A^{-1/6}
\end{eqnarray}
(A is the atomic mass of the nucleus).
Even though this formula refers to pp- and nn-reactions, it can however be used to our results referred to pn-reactions ($\mu^-$-capture), on the basis of the well known Foldy-Walecka theorem according to which the giant dipole resonance in $\mu^-$-capture rates, are calculated starting from the experimental photo-absorption cross sections \cite{Foldy-Wal-64}.
According to Eq. (\ref{giant-dipole-resonance}) for $^{48}Ti$ the maximum $1^{-}$ peak is located at 18.668 MeV, for $^{56}Fe$ at 18.716 MeV, for $^{66}Zn$ at 17.945 MeV and for $^{90}Zr$ at 16.684 MeV which are in a good agreement with our results (the worst case occurs for  $^{66}Zn$ where the empirical peak is at about 15 MeV).
Moreover, our results are in good agreement with the conclusions of Ref. \cite{Eram-Kuz-Tet-98}
where authors mention that for the stable Ni isotopes ($^{58.60,62}Ni$) the peak appears in the range of 18-19 MeV.
We note that similar conclusion is extracted from the study of the charged current reaction $^{56}Fe(\nu_{e},e^{-})^{56}Co$ by Kolbe and Langanke \cite{Kol-Lang-99}, where the peak of the giant dipole resonance appears at about 17 MeV (see Fig 1 of Ref. \cite{Kol-Lang-99}).

As can be seen from Figs. \ref{Si28-S32-Sort}, \ref{Ti48-Fe56-Sort} and \ref{Zn66-Zr90-Sort}, the main contributions coming from the polar-vector operator are the $1^{-}$ and $0^{+}$ states while, the most important transitions due to the axial-vector operator are the $0^{-}$, $1^{+}$ and $2^{-}$ excitations, namely the lowest spin states.

We note that the figures of this section, have been designed by using the ROOT program of Cern with binning width 0.112, 0.105, 0.105, 0.15, 0.14, 0.11, respectively, for $^{28}Si$, $^{32}S$, $^{48}Ti$, $^{56}Fe$, $^{66}Zn$, $^{90}Zr$ nuclei.

\subsection{Contribution of Multipole Transitions}
\label{Mult-Trans}

The second step of our study includes calculations of the partial $\mu^{-}$-capture rates for various low-spin multipolarities, $\Lambda_{J^{\pi}}$ (for $J^{\pi}\leq 4^{\pm}$), in the chosen set of nuclei. These partial rates have been found by summing over the contibutions of all the individual multipole states of the studied multipolarity as
\begin{eqnarray}
\Lambda_{J^{\pi}} & = & \sum_{f} \Lambda_{gs \rightarrow J_{f}^{\pi}} = 2G^{2}\langle \Phi _{1s}  \rangle^{2} \cdot\\ 
&\Big[& \sum_{f} q_{f}^{2} R_{f} \big|\langle J_{f} ^{\pi}\Vert (\widehat{\mathcal{M}}_{J}-\widehat{\mathcal{L}}_{J})\Vert 0_{gs}^{+}\rangle \big|^{2}\nonumber\\
&+&\sum_{f} q_{f}^{2} R_{f}\big|\langle J_{f} ^{\pi}\Vert (\widehat{\mathcal{T}}_{J}^{el}-\widehat{\mathcal{T}}_{J}^{magn})\Vert 0_{gs}^{+}\rangle \big|^{2} \Big]\nonumber
\end{eqnarray}
where $f$ runs over all states of the multipolarity $\vert J^{\pi}\rangle$.
As mentioned before, these calculations have been performed first by using the free nucleon axial-vector coupling constant $g_{A} = 1.262$, and then by taking into account the quenching effect indicated for medium-weight nuclei with $g_{A} = 1.135$. 

\begin{figure*}
\begin{center}
\includegraphics[scale = 1.00]{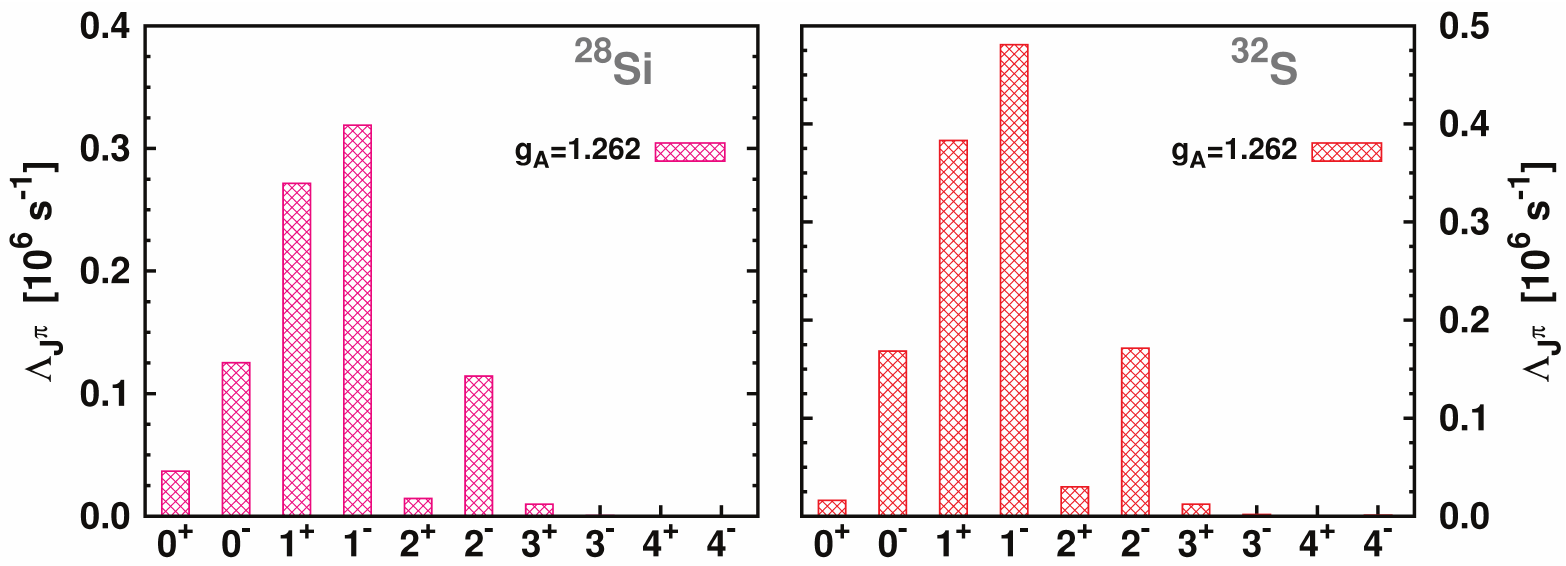}
\end{center}
\vspace{-0.7 cm }
\caption{\label{Figure S-muc-rates} Partial muon capture rates $\Lambda_{J^{\pi}}$ of different multipole transitions in $^{28}Si$ and $^{32}S$ isotopes. In both isotopes the pronounced contributions are the $J^{\pi}=1^{-}$ and $J^{\pi}=1^{+}$ multipolarity.}
\end{figure*}

For the target $^{28}Si$ \cite{Mof-Mes-97,Brud-Egor-95} and $^{32}S$ isotopes, these calculations were performed only for the free nucleon coupling constant $g_{A} = 1.262$. (the quenching  effect can be ignored \cite{Kuz-Tet-02}). The results obtained for the partial $\mu^{-}$-capture rates of these isotopes are illustrated in Fig. \ref{Figure S-muc-rates},
from which one can see that, as it is expected, the most important multipole transitions are the $J^{\pi} = 1^{+}$ and $1^{-}$.
More specifically, for $^{28}Si$ isotope, the contributions of all $J^{\pi} = 1^{-}$ transitions exhaust the $36\%$ of the total muon-capture rate and the $J^{\pi} = 1^{+}$ about $30\%$. Significant contribution, about $14\%$, comes from the $J^{\pi} = 0^{-}$ multipolarity and about $13\%$ from the $J^{\pi} = 2^{-}$.
A similar picture is found in $^{32}S$ isotope, where 
the dominant contributions to the total muon-capture rate are the $J^{\pi} = 1^{-}$ ($38\%$) and the $J^{\pi} = 1^{+}$ ($30\%$).
From the rest of the multipolarities rather significant portions come from the abnormal parity transitions $0^{-}$ and $2^{-}$ about $13 \%$ and $14 \%$ respectively. 


Because, as mentioned in the Indroduction, for electromagnetic and weak charged-current nuclear processes, the free nucleon coupling constant $g_{A}$ must be modified for medium-weight and heavy nuclei \cite{Zin-Lang-06},
in $\mu^{-}$-capture on $^{48}Ti, ^{56}Fe, ^{66}Zn, ^{90}Zr$ isotopes we repeated the state-by-state calculations by using $g_{A}=1.135$ (a value smaller by about $10-12 \%$ compared to the $g_{A} = 1.262$).
Historically, the necessity of the renormalization of $g_{A}$, came out of the following studies:
(i) In the analysis of measurements on the nuclear beta-decays that lead to low-lying excitations \cite{Wild-84}, and 
(ii) in the interpretation of the missing Gamow-Teller strength revealed in forward angle (p,n) and (n,p) charge-exchange reactions \cite{Hau-91}. 
We note that, in (n,p) reactions many authors use quenched values of $g_A$ lying in the region of $0.9 < g_A < 1.0$ for nuclei with mass number $41 < A < 64$ \cite{Faess-09,Lang-Dean-Rad-95,Pin-Po-96}.
In $\beta^{-}$-decay and (p,n) reactions the quenching is mainly related to the neglect of configurations outside the model space used and the non-consideration of the meson-exchange currents.
 
A quenched value of $g_{A}$ was recently suggested to be used in other weak interaction processes such as the neutrino induced nuclear reactions.
As has been found \cite{Kuz-Tet-02}, the consideration of a quenched factor instead of the free nucleon axial-vector coupling constant, leads to better agreement of the theoretical results with the experimental muon capture rates.
Since the axial-vector form factor $F_{A}(q^{2})$ multiplies all four operators (see Eqs. (\ref{Coul})-(\ref{Trans-Magn})), a quenched value of $g_{A}$ must enter the multipole operators generating the pronounced excitations $0^{-},1^{\pm}...$ etc. In Ref. \cite{Zin-Lang-06}, a quenched value of $g_{A}$ is used only for the true Gamow-Teller transitions. 
In our study, we find that for the reproducibility of the experimental data, as the mass number A of the nucleus increases the quenching becomes more signifficant and can not be ignored as we have done in the case of the $^{28}Si$ and $^{32}S$ isotope. 

\begin{figure*}
\begin{center}
\includegraphics[scale = 1.00]{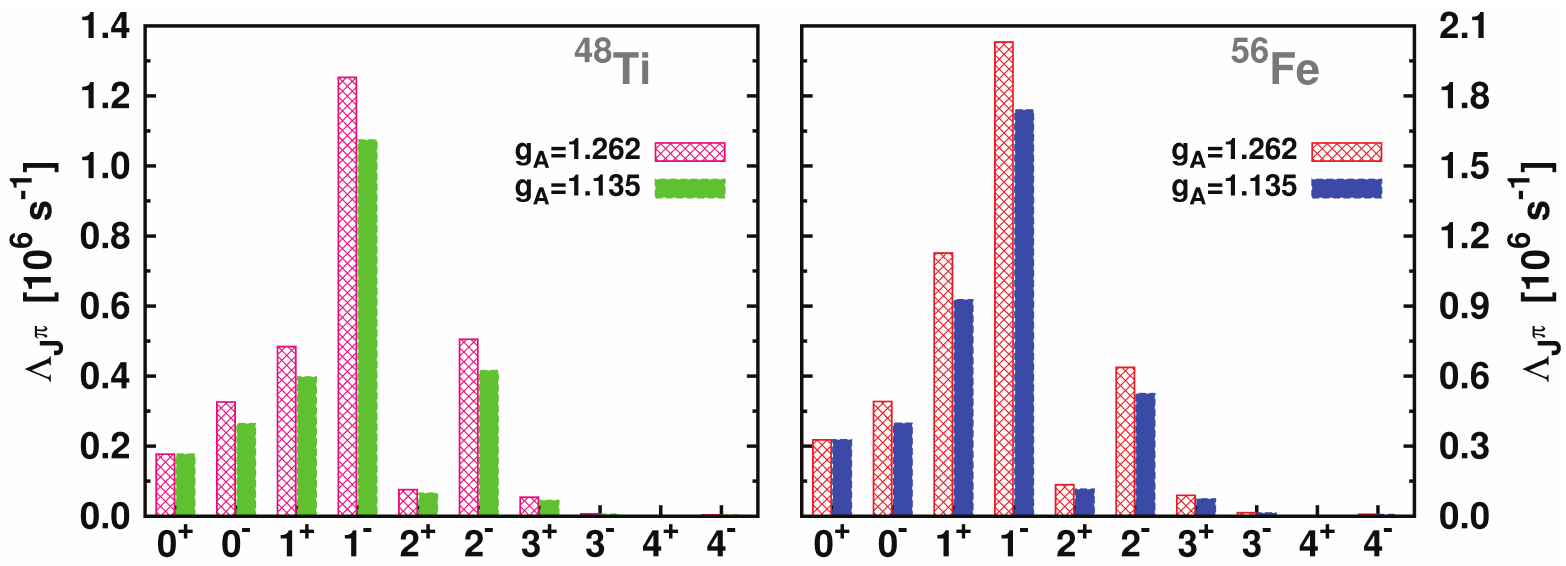}
\includegraphics[scale = 1.00]{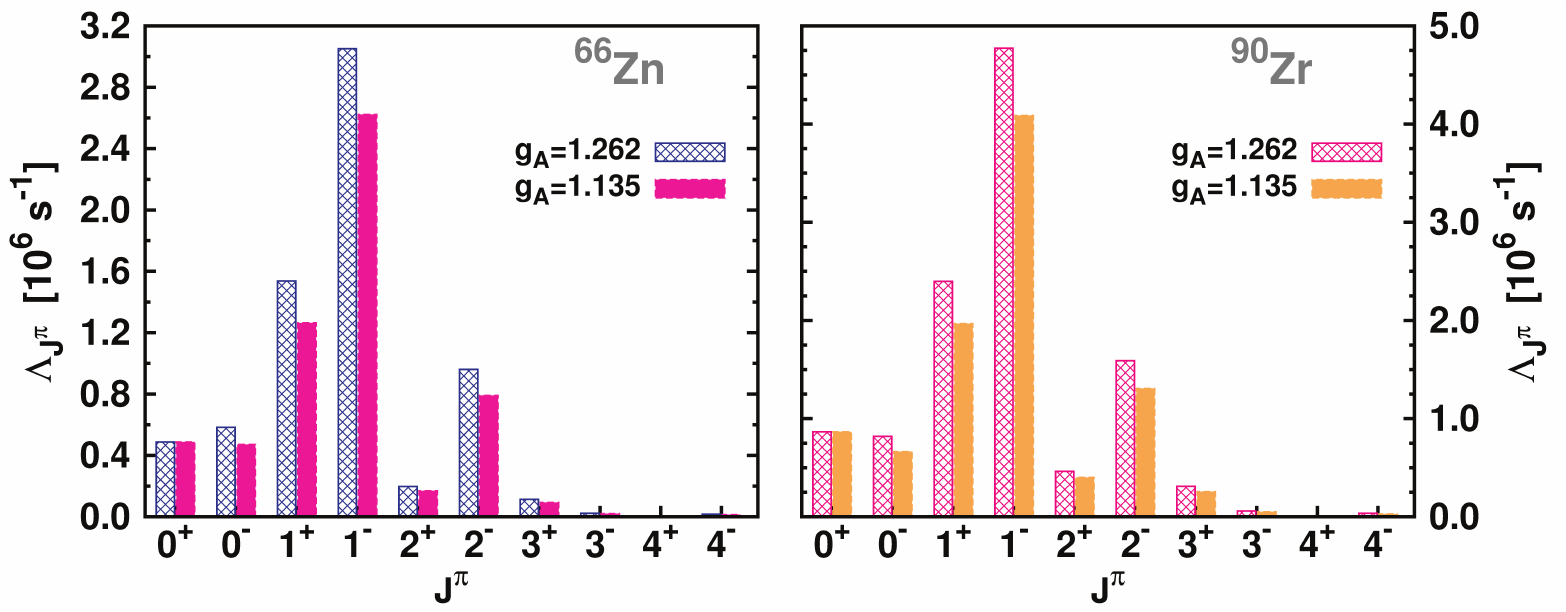}
\end{center}
\vspace{-0.5 cm }
\caption{\label{Ti-Fe-Zn-Zr-mucap-rates} Contribution of multipole transition rates $\Lambda_{J^{\pi}}$ (up to $J^{\pi} = 4^{\pm}$) with the total muon capture rate in $^{48}Ti$, $^{56}Fe$,  $^{66}Zn$ and $^{90}Zr$ isotopes with (filled histograms) and without (double dashed histogramms) quenching effect. The dominance of $J^{\pi}=1^{-}$ and $1^{+}$ multipolarities is obvious in all nuclei.}
\end{figure*}

For the medium-weight nuclei $^{48}Ti, ^{56}Fe, ^{66}Zn$ and $^{90}Zr$, we used the moderate quenched value  $g_{A} = 1.135$ and found that our rates are in good agreement with the results of other works \cite{Zin-Lang-06}. 
By using this value of $g_{A}$ for the contributions of the different multipole transitions in the isotopes $^{56}Fe, ^{66}Zn$ and $ ^{90}Zr$, we found that the most important peaks correspond to the  $J^{\pi} = 1^{+}$ and $1^{-}$. For the $^{48}Ti$ isotope, however, we found that a great part of the total rate comes from the $J^{\pi} = 1^{-}$ and $2^{-}$, as is shown in Fig. \ref{Ti-Fe-Zn-Zr-mucap-rates}.
In more detail, in the case of $^{48}Ti$ isotope the $1^{-}$ multipolarity contributes about $44 \%$, the $2^{-}$ about $17 \%$, the $1^{+}$ about $16 \%$ and the $0^{-}$ about $11 \%$. Significant contribution (about $7 \%$) originates also from the $0^{+}$ multipolarity.
For $^{56}Fe$ isotope the most important contribution about $42 \%$ comes from the $1^{-}$ multipolarity. Other multipolarities with significant contributions are the $1^{+} (22 \%), 2^{-} (13 \%), 0^{-} (10 \%)$ and $0^{+} (8 \%)$. A similar picture appears in the other two isotopes, $^{66}Zn$ and  $^{90}Zr$, where the major contribution is derived from the $1^{-}$ multipolarity, about $44 \%$  and $42 \%$ respectevelly. The $1^{+}$ multipolarity contributes about $21 \%$ in  $^{66}Zn$ and  about $20 \%$ in $^{90}Zr$ isotope.
Correspondingly, the $2^{-}$ contributes about $13 \%$ for $^{66}Zn$ and about $14 \%$ for $^{90}Zr$, the $0^{+}$ about $8 \%$ and $9 \%$ respectivelly and finally the $0^{-}$ multipolarity offers about $8 \%$ for $^{66}Zn$ and about $7 \%$ for $^{90}Zr$.

In Table \ref{mult-contr} we present the partial muon-capture rates obtained for the low-spin multipole transitions up to $J^{\pi} = 4^{\pm}$ evaluated with our pn-QRPA code. Correspondingly, in Table \ref{percentage-of total mucap} we tabulate the individual portions to the total OMC rate, for the low-spin nultipole transitions up to $ J^{\pi}=4^{\pm}$. As can be seen, for all nuclei the contribution of $1^{-}$ multipole transitions is the most important multipolarity, exhausting more than $39\%$ of the total muon-capture rate. Ordinary muon capture proceeds mainly through spin-multipole transitions, the most important of which are the Gamow-Teller transitions ($j_{0}(kr)\sigma t^{+}$ operator), and the spin-dipole transitions ($j_{1}(kr)[Y_{1}\otimes\sigma]^{J}t^{+}$ operator) where $j_{0}$ and $j_{1}$ are the spherical Bessel functions of zero and first order, respectively \cite{Eram-Kuz-Tet-98}. Such important contribution is found in $^{16}O$ and in $^{48}Ca$ isotopes studied in Ref. \cite{Kol-Lang-00}.

There are no similar results for the isotopes  $^{28}Si$, $^{32}S$, $^{48}Ti$, $^{56}Fe$ and  $^{66}Zn$ to compare with our portions. For the $^{90}Zr$, however, Kolbe, Langanke and Vogel \cite{Kol-Lang-00} found about $28 \%$ (for $1^{-}$), $25 \%$ (for $1^{+}$) and about $13 \%$ (for $2^{-}$) multipolarities which, with the exception of $1^{-}$ contribution, are in good agreement with our results listed in Table \ref{percentage-of total mucap}. The difference in $1^{-}$ multipolarity is mostly due to the fact that $^{90}Zr$ is a double closed shell nucleus and the QRPA convergence is treated as in Ref. \cite{Kos-Faes-97,Suh-93}.

\begin{table}[t]
 \caption{\label{mult-contr} Muon capture rates $\Lambda_{J^{\pi}}$ (in $10^{6}\,s^{-1}$) of each multipolarity evaluated with our pn-QRPA code.}
 \begin{center}
\begin{tabular}{l| lllllll }
 \hline
 \hline\\[-0.3cm]
 &  $^{28}Si$  &  $^{32}S$  &  $^{48}Ti$  & $^{56}Fe$ &  $^{66}Zn$ &$^{90}Zr$ \\
 \hline
 $0^{-}$ & 0.125 & 0.168 & 0.264 & 0.398 & 0.471 & 0.662 \\[0.2ex]
 \hline
$0^{+}$ &  0.037 & 0.016 & 0.177  & 0.327  & 0.488  & 0.866 \\[0.2ex]
 \hline
 $1^{-}$ & 0.319 & 0.481 & 1.074 & 1.740 & 2.623 & 4.087 \\[0.2ex]
  \hline
 $1^{+}$ & 0.271 & 0.383 & 0.397 & 0.926 & 1.263 & 1.968 \\[0.2ex]
 \hline
 $2^{-}$ & 0.114 & 0.171 & 0.415 & 0.524 & 0.790 & 1.307 \\[0.2ex]
  \hline
 $2^{+}$ & 0.014 & 0.030 & 0.065 & 0.115 & 0.169  & 0.401  \\[0.2ex]
 \hline
 $3^{-}$ & 0.001 & 0.002 & 0.006 & 0.013 & 0.020 & 0.050 \\[0.2ex]
  \hline
 $3^{+}$ & 0.010 & 0.012 & 0.045 & 0.073  & 0.093 & 0.255 \\[0.2ex]
 \hline
 $4^{-}$ & 0.001 & 0.001 & 0.003 & 0.006 & 0.014 & 0.029  \\[0.2ex]
 \hline
 $4^{+}$ & $0.2\, 10^{-4}$ & $0.8\, 10^{-4}$ & $0.2\, 10^{-3}$ &$0.5\, 10^{-3}$  & $0.7\,  10^{-3}$  & $2.5\, 10^{-3}$ \\[0.2ex]
 \hline
\hline
\end{tabular}
\end{center}
\end{table}

\begin{table}[t]
 \caption{\label{percentage-of total mucap} The percentage of each multipolarity into the total muon-capture rate evaluated with our pn-QRPA code.}
 \begin{center}
 \begin{tabular}{ R{0.8cm}| R{1.2cm} R{1.2cm} R{1.2cm}R{1.2cm}R{1.2cm}R{1.2cm}}
 \hline
 \hline\\[-0.3cm]
  &  $^{28}Si$  &  $^{32}S$  &  $^{48}Ti$  & $^{56}Fe$ &  $^{66}Zn$ &$^{90}Zr$ \\
 \hline
 $0^{-}$ & 14.03 & 13.30 & 10.78 & 9.64 & 7.94  & 6.89 \\[0.2ex]
 \hline
 $0^{+}$ &  4.11 & 1.27 & 7.24  & 7.92 & 8.22 & 8.99 \\[0.2ex]
 \hline
 $1^{-}$ & 35.74 & 38.01 & 43.88 & 42.18 & 44.21 & 42.43 \\[0.2ex]
 \hline
 $1^{+}$ & 30.42 & 30.28 & 16.24 & 22.46 & 21.29 & 20.43 \\[0.2ex]
 \hline
 $2^{-}$ & 12.81 & 13.54 & 16.97 & 12.72 & 13.32 & 13.57 \\[0.2ex]
 \hline
 $2^{+}$ &  1.62 & 2.36 & 2.67 & 2.79 & 2.85 & 4.16 \\[0.2ex]
 \hline
 $3^{-}$ &  0.10  & 0.15 & 0.23  & 0.32  & 0.34 & 0.52 \\[0.2ex]
 \hline
 $3^{+}$ &  1.09 & 0.97 & 1.82  & 1.78 & 1.58 & 2.65 \\[0.2ex]
 \hline
 $4^{-}$ &  0.06 & 0.10 & 0.14 & 0.16 & 0.23 & 0.30 \\[0.2ex]
 \hline
 $4^{+}$ &  0.01 & 0.01 & 0.01 & 0.01  & 0.01  & 0.03\\[0.2ex]
 \hline
\hline
\end{tabular}
\end{center}
\end{table}

\subsection{Total Muon-Capture-Rates}
\label{Total Muon Capture Rates}
In the last stage of our present work, we computed the total rates of muon-capture on the chosen set of nuclei. These rates are obtained by summing over all partial multipole transition rates in two steps. At first, we sum up the contribution of each final state of a specific multipolarity, and then, we sum over the multipole responses (up to $J^{\pi} = 4^{\pm}$) as
\begin{eqnarray}
\Lambda_{tot} = \sum_{J^{\pi}} \Lambda_{J^{\pi}} = \sum_{J^{\pi}} \sum_{f} \Lambda_{J_{f}^{\pi}}
\end{eqnarray}
Such calculations have been carried out twice: one with $g_{A} = 1.262$ (free nucleon axial-vector coupling constant) and the other with the quenched value $g_{A}= 1.135$ \cite{Zin-Lang-06}. The results are tabulated in Table \ref{Tot-mu-cap}, where for the sake of comparisson we also include the experimental total rates as well as the theoretical ones of Ref. \cite{Zin-Lang-06}. Moreover, in Table \ref{Tot-mu-cap} we show the individual contribution in the total muon capture rate of the polar-vector ($\Lambda^{V}_{tot}$), the axial-vector ($\Lambda^{A}_{tot}$), and the overlap ($\Lambda^{VA}_{tot}$) parts.

 \begin{table}[h]
 \caption{Individual contribution of Polar-vector, Axial-vector and Overlap part to the total muon-capture rate. Comparison between the total muon capture rates obtained by using the pn-QRPA with the quenched value of $g_{A}= 1.135$ for medium-weight nucleus ($^{48}Ti, ^{56}Fe, ^{66}Zn$ and $^{90}Zr$) and the free nucleon coupling constant $g_{A}= 1.262$ for the light nucleus $^{28}Si$ and $^{32}S$, with the available experimental data and with the theoretical rates of Ref \cite{Zin-Lang-06}.}
 \label{Tot-mu-cap}
 \begin{small}
\begin{tabular}{c| c c c c c c  }
 \hline
 \hline
 \multicolumn{7}{c}{ Total Muon-capture rates $\Lambda_{tot} (\times 10^{6}) s^{-1}$} \\[0.5ex]

 \cline{1-7}\\[-0.3cm]
& \multicolumn{4}{c}{pn-QRPA Calculations} & Experiment & RPA Dean\\[0.5ex]
 \cline{1-7}\\[-0.3cm]
   Nucleus &  $\Lambda^{V}_{tot}$  & $\Lambda^{A}_{tot}$ & $\Lambda^{VA}_{tot}$ & $\Lambda_{tot}$ & $\Lambda^{exp}_{tot}$ & $\Lambda^{theor}_{tot}$ \cite{Zin-Lang-06} \\ [0.5ex]

  \hline \\[-0.3cm]
    $^{28}Si$ & 0.150 & 0.751 & -0.009  & 0.892 & 0.871 & 0.823  \\[0.5ex]
    $^{32}S $ & 0.204 & 1.078 & -0.017  & 1.265 & 1.352 & 1.269  \\[0.5ex]
    $^{48}Ti$ & 0.628 & 1.902 & -0.081  & 2.447 & 2.590 & 2.214  \\[0.5ex]
    $^{56}Fe$ & 1.075 & 3.179 & -0.129  & 4.125 & 4.411 & 4.457  \\[0.5ex]
    $^{66}Zn$ & 1.651 & 4.487 & -0.204  & 5.934 & 5.809 & 4.976   \\[0.5ex]
    $^{90}Zr$ & 2.679 & 7.310 & -0.357  & 9.631 & 9.350 & 8.974   \\[0.5ex]

\hline
\hline
\end{tabular}
\end{small}
\end{table}

\begin{figure}[h!]
\begin{center}
\includegraphics[scale = 1.2]{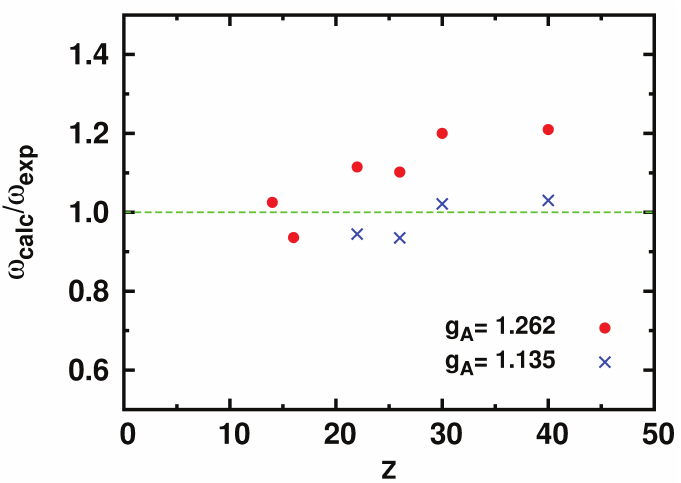}
\end{center}
\caption{\label{ratio} Ratio of the calculated and experimental total muon capture rates as a function of Z. Circles and X symbols correspond to rates calculated with free nucleon $g_{A}$ and quenched value of $g_{A}$ respectively.}
\end{figure}

As can be seen, our results obtained with the quenched $g_{A}$ are in very good agreement with the experimental total muon-capture rates.  For all studied nuclei the deviations from the corresponding experimental rates are smaller than $7\%$ when using the quenched $g_{A}$ (the deviation is much bigger when using the  $g_{A} = 1.262$). So, for the reliability of our results it is necessary to take into account the quenching effect. 
To make it more perceptual, in Fig. \ref{ratio} we have plotted the ratio of our theoretical total muon-capture rates divited by the experimental ones, i.e.
\begin{eqnarray}
\lambda = \frac{\omega_{calc}}{\omega_{exp}}
\end{eqnarray}
for the results obtained with the above two values of $g_{A}$ (with and without quenching). The filled circles represent the results for the free $g_{A}$ and the X symbols the results for the quenched $g_{A}$.
It is evident the better agreement of our calculations with quenched value of $g_{A}$.
We furthermore, compare our results with the available calculated rates Zinner \cite{Zin-Lang-06} obtained by using different approach and the comparison is good.

Finally, it is worth noticing that, in medium-weight nuclei the contribution comes mainly from transitions for which the angular momentum transfer is L=0,1 and 2 but, in heavy nuclei, some contributions from higher multipolarities become noticeable.

\section{Summary and Conclusion}
In the present work, relying on an advantageous numerical approach constructed by our group recently, we performed detailed calculations for all
multipole transition matrix elements entering the exclusive muon-capture
rates. The required nuclear wave functions were obtained
within the context of the pn-QRPA using realistic two-body forces (Bonn C-D potential). Results for the exclusive rates through extensive state-by-state calculations and subsequently for the total muon capture rates on the set of isotopes $^{28}Si$, $^{32}S$, $^{48}Ti$, $^{56}Fe$, $^{66}Zn$ and $^{90}Zr$ were computed. 

Because the capture rates are rather sensitive to the quenching of the axial-vector coupling constant, we examined  the known as in-medium effect of the nucleon, by reducing this constant from its free nucleon value $g_{A} = 1.262$ to the effective value $g_{A} = 1.135$ for all multipole transitions, and found that the experimental muon capture rates are well reproduced with an accuracy better than $10\%$.
Detailed study of this effect, however, required for experiments at RCNP \cite{Kos-Ejiri,kos-14} is under way and results are expected to be obtained soon.

The muon-capture studies on these nuclei demonstrate that the used pn-QRPA method may provide an accurate description of the charged current semileptonic weak interaction processes in the Z-range of the isotopes chosen. As the inclusive muon capture rates and the cross section of the antineutrino-induced charged current reactions are closely related (both of them are governed by the same nuclear matrix elements and proceed via the same initial and final states), we have adopted this method to study other types of charge-changing weak interaction processes as, electron-capture, beta-decay modes, etc. \cite{Giannaka,Gian-Kosm-15} in currently interesting nuclei from a nuclear astrophysics point of view.

 \subsection*{Acknowledgments}
 This research has been co-financed by the European Union (European Social Fund-ESF) and Greek national funds through the Operational Program ``Education and Lifelong Learning" of the National Strategic Reference Framework (NSRF) - Research Funding Program: Heracleitus II. Investing in knowledge society through the European Social Fund.
 \appendix
\section{Nuclear Matrix Elements}
 \label{Nuclear Matrix Elements}
The eight different tensor multipole operators entering  Eq. (\ref{muon-cap-rates}) refer to as Coulomb $\widehat{\mathcal{M}}_{J}$, longitudinal $\widehat{\mathcal{L}}_{J}$, transverse electric $\widehat{\mathcal{T}}^{el}_{J}$ and transverse magnetic  $\widehat{\mathcal{T}}^{magn}_{J}$, contain polar-vector as well as axial-vector parts and are written as:
\begin{flushleft}
\begin{eqnarray}
\label{Coul}
\hspace{-3.0 cm}
\widehat{\mathcal{M}}_{JM}(qr) &=& \widehat{M}^{coul}_{JM} + \widehat{M}^{coul5}_{JM}\\
&=& F_{1}^{V} M^{J}_{M} (qr) - i\frac{q}{M_{N}}[F_{A}\Omega^{J}_{M}(qr)\nonumber\\
&+& \frac{1}{2}(F_{A}+q_{0}F_{p})\Sigma^{''J}_{M}(qr)]\nonumber \\
\label{Long}
\widehat{\mathcal{L}}_{JM}(qr) &=& \widehat{L}_{JM} + \widehat{L}^{5}_{JM}\\
&=& \frac{q_{0}}{q}F_{1}^{V} M^{J}_{M} (qr) + iF_{A}\Sigma^{''J}_{M}(qr)\nonumber\\
\label{Trans-Electr}
\widehat{\mathcal{T}}^{el}_{JM}(qr) &=& \widehat{T}^{el}_{JM} + \widehat{T}^{el5}_{JM}\\
&=& \frac{q}{M_{N}}[F_{1}^{V} \Delta^{'J}_{M} (qr) + \frac{1}{2}\mu^{V} \Sigma_{M}^{J}(qr)] \nonumber\\
&+& iF_{A}\Sigma^{'J}_{M}(qr)\nonumber\\
\label{Trans-Magn}
\widehat{\mathcal{T}}^{magn}_{JM}(qr) &=& \widehat{T}^{magn}_{JM} + \widehat{T}^{magn5}_{JM}\nonumber\\
&=& -\frac{q}{M_{N}}[F_{1}^{V} \Delta^{J}_{M} (qr) \nonumber\\
&-& \frac{1}{2}\mu^{V} \Sigma^{'J}_{M}(qr)] + iF_{A}\Sigma^{J}_{M}(qr)
\end{eqnarray}
\end{flushleft} 
where the form factors $F_{X}$, X=1,A,P and $\mu^{V}$ are functions of the 4-momentum transfer $q^{2}_{\mu}$.

These multipole operators, due to the Conserved Vector Current (CVC) theory, are reduced to seven new basic operators expressed in terms of spherical Bessel functions, spherical harmonics and vector spherical harmonics (see Refs. \cite{DonPe,Has4}). 
The single particle reduced matrix elements of the form $\langle j_{1} \Vert T_{i}^{J}\Vert j_{2}\rangle$, where $T_{i}^{J}$ represents any of the seven basic multipole operators ($M_{M}^{J}$, $\Omega_{M}^{J}$, $\Sigma_{M}^{J}$, $\Sigma_{M}^{'J}$, $\Sigma_{M}^{''J}$, $\Delta_{M}^{J}$, $\Delta_{M}^{'J}$) of Eq. (\ref{Coul})-(\ref{Trans-Magn}), have been written in closed compact formulae as \cite{Has4,Ts-Kos-11}
\begin{eqnarray}
\langle (n_{1}l_{1})j_{1} \Vert T^{J}\Vert (n_{2}l_{2})j_{2}\rangle = e^{-y} y^{\beta/2}\sum_{\mu=0}^{n_{max}}P_{\mu}^{J}y^{\mu}
\end{eqnarray}
where the coefficients $P_{\mu}^{J}$ are given in Ref. \cite{Has4}.
In the latter summation the upper index $n_{max}$ represents the maximun harmonic oscillator quanta included in the active model space chosen as $n_{max} = (N_1+N_2-\beta)/2$, where $N_{i} = 2n_{i}+l_{i}$, i=1,2, and $\beta$  is related to the rank of the above operators \cite{Has4}.

In the context of the pn-QRPA, the required reduced nuclear matrix element between the initial $\vert 0^{+}_{gs}\rangle$ 
and any final $\vert f\rangle$ state entering the rates of Eq. (\ref{OMC-rates}) are given by
\begin{eqnarray}
\label{red-matr-elem}
\langle f \Vert\widehat{T}^{J}\Vert 0^{+}_{gs} \rangle &=& \sum _{j_{2}\geq j_{1}} 
\frac{\langle j_{2}\Vert\widehat{T}^{J}\Vert j_{1} \rangle}{[J]}  \nonumber\\
&\cdot & \left[ X_{j_{2}j_{1}} u_{j_{2}}^{p} \upsilon_{j_{1}}^{n} + Y_{j_{2}j_{1}} \upsilon_{j_{2}}^{p} u_{j_{1}}^{n} \right]
\end{eqnarray}
where $u_j$ and $\upsilon_j$ are the probability amplitudes for the $j$-level to be unoccupied or occupied, respectively (see the text) \cite{Kos-Verd-94}.

These matrix elements enter the description of various semi-leptonic weak interaction processes in the presence of nuclei \cite{Don-Wal-72,Don-Wal-73,Con-Don-72,Don-Wal-76,DonPe,Has4,Gian-Kos-13,Ts-Kos-11,Bal-Ydr-11,Balasi-Ydr-11,tsak-kos-11,Bal-Ydr-12,kos-ts-12,Pap-Kos-14,Pap-Kos-13}.

\section{Muon wave function in the muonic atom} 
\label{Wave function}

The calculation of the exact muon wave function, $\Phi_{1s}(\textbf{r})$, entering Eq. (\ref{muon-cap-rates}) needs the use of a specific numerical method. This however, can be avoided by using either its value at $r \simeq 0$, namely the  $\Phi_{1s}(r \simeq 0)$, or as stated in Sect. \ref{Formalism of muon capture rates}, an average value $\langle \Phi_{1s} \rangle $, which is  given in terms of the effective nuclear charge $Z_{eff}$ that sees the bound muon as 
\begin{eqnarray}
\langle\Phi_{1s}\rangle ^{2} = \frac{1}{\pi}\alpha^{3}m_{\mu}^{3}\frac{Z_{eff}^{4}}{Z}
\end{eqnarray}
($\alpha$ denotes the fine structure constant). The quantity $Z_{eff}$ is approximated by 
$Z_{eff}^{4} = \pi \alpha_{0}^{3} \langle \rho\rangle$,
where $\alpha_{0}$ is the muon Bohr radius and $\langle \rho\rangle$ is the mean charge density of the parent nucleus \cite{Ford-Wills-62}.\, For light nuclei $Z_{eff}\simeq Z$ but for heavier ones $Z_{eff}\ll Z$.
In recent studies the exact wave functions for the bound muon are obtained by solving the Schroedinger and Dirac equations by using neural network techniques or genetic algorithms \cite{Kos-Lag-02}. In the work of Zinner, Langanke and Vogel \cite{Zin-Lang-06}, for the description of the exact bound muon wave functions (w-fs), the muon density beyond the site of the nucleus is considered for solving the Dirac equation. These authors use exact muon wave functions, for other muonic orbits, $\Phi_{2p}$, etc, which are considered to have rather signifficant contributions \cite{Zin-Lang-06}.

\section{pn-QRPA equations}
\label{pn-QRPA equations}
  In our numerical solution performance we rewrite the QRPA equations (\ref{matr-form-QRPA}) by defining a new set of amplitudes $P^{m}$ and $R^{m}$ which are related to the proceding ones through 
 \begin{eqnarray}
 X^{m} &=& \sqrt{\frac{1}{2}}(\Omega_{m}^{1/2}P^{m} + \Omega_{m}^{-1/2}R^{m})\nonumber\\
 Y^{m} &=& \sqrt{\frac{1}{2}}(-\Omega_{m}^{1/2}P^{m} + \Omega_{m}^{-1/2}R^{m})
 \end{eqnarray}
 The new amplitudes satisfy the matrix expressions 
  \begin{equation}
 \mathcal{(A-B)}P^{m} =  R^{m}\, , \quad
 \mathcal{(A+B)}R^{m} = \Omega^{2}_{m} P^{m}
 \end{equation}
 Then, we have
 \begin{eqnarray}
 \mathcal{(A+B)(A-B)}P^{m} = \Omega ^{2}_{m} P^{m}.
 \end{eqnarray} 
The latter equations can be diagonalized seperately and, subsequently, the X and Y amplitudes are directly determined \cite{Kam-Faes-91,Bar-60}. 


\bibliographystyle{model6-num-names}
\bibliography{<your-bib-database>}



\end{document}